\documentclass[aps,prd,twocolumn,groupedaddress,showpacs]{revtex4-1}
\usepackage{epsf,epsfig,graphicx}
\usepackage{amsmath}
\usepackage{amssymb}
\usepackage{tikz}

\def\edth{\;\raise1.0pt\hbox{$'$}\hskip-6pt\partial\;}
\def\baredth{\;\overline{\raise1.0pt\hbox{$'$}\hskip-6pt
\partial}\;}
\def\gsim{~\rlap{$>$}{\lower 1.0ex\hbox{$\sim$}}}

\newcommand{\be}{\begin{equation}}
\newcommand{\ee}{\end{equation}}
\newcommand{\bw}{\begin{widetext}}
\newcommand{\ew}{\end{widetext}}

\newcommand{\intinf}{\int_{-\infty}^{\infty}}
\newcommand{\suml}{\sum_{l=0}^{\infty}}
\newcommand{\summ}{\sum_{m=-\ell}^{\ell}}

\usepackage{xcolor}

\usepackage[colorlinks]{hyperref}
\definecolor{darkblue}{HTML}{2E3092}
\hypersetup{colorlinks=true,linkcolor=darkblue,citecolor=darkblue,urlcolor=darkblue}

\AtBeginDocument{\renewcommand{\d}{\mathbf{d}}}

\begin{document}

\title{Spherical harmonic analysis of anisotropies in polarized stochastic gravitational wave
background with interferometry experiments}

\author{Yu-Kuang Chu$^{1}$, Guo-Chin Liu$^{2}$, and Kin-Wang Ng$^{1,3}$}

\affiliation{
$^1$Institute of Physics, Academia Sinica, Taipei 11529, Taiwan\\
$^2$Department of Physics, Tamkang University, Tamsui, New Taipei City 25137, Taiwan\\
$^3$Institute of Astronomy and Astrophysics, Academia Sinica, Taipei 11529, Taiwan
}

\vspace*{0.6 cm}
\date{\today}
\vspace*{1.2 cm}

\begin{abstract}
We study the interferometric observation of intensity and polarization anisotropies of a stochastic gravitational wave background (SGWB). We show that the observed correlated data is defined in the group manifold of the three-dimensional rotation. Explicit correlation between two detectors in the interferometry experiments such as LIGO-Virgo and KAGRA is constructed in terms of the Wigner D-functions. Our results may provide a tool for constructing data pipelines to estimate the power spectra of the SGWB anisotropies.
\end{abstract}

\maketitle

\section{Introduction}

The LIGO detectors have firstly observed gravitational waves (GWs) emitted by a binary black hole merger as predicted in general relativity~\cite{ligo}. 
Since then, a handful of GW events from compact binary coalescences has been observed in Advanced LIGO and Advanced Virgo O2 and O3 observing runs~\cite{ligo2019}. This achievement has opened up a new era of GW astronomy and cosmology. Future experimental plans such as Einstein Telescope~\cite{et2010}, Cosmic Explorer~\cite{ce2019}, LISA~\cite{lisa2017}, DECIGO~\cite{decigo2011}, Taiji~\cite{taiji2017}, Tianqin~\cite{tianqin2016}, and pulsar-timing arrays like SKA~\cite{ska2015} will bring us a precision science in GW observation~\cite{ligo2050}.

Stochastic gravitational wave background (SGWB) is a key target in GW experiments. There have been many studies on possible astrophysical and cosmological sources for SGWB such as distant compact binary coalescences, early-time phase transitions, cosmic string or defect networks, second-order primordial scalar perturbations, and inflationary GWs~\cite{romano}. GWs have very weak gravitational interaction, so they decouple from matter at the time of production and then travel to us almost without being perturbed. At the present, they remain as a SGWB that carries original information of the process of production in the very early universe. 

In general, the SGWB can be anisotropic and polarized. For examples, helical GWs can be produced in axion inflation models, leading to a net circular polarization~\cite{alexander,satoh,sorbo,crowder2013}. Linear polarization can be generated through diffusion by compact astrophysical objects, with an amount suppressed by a factor of at least $10^{-4}$ with respect to the intensity anisotropies; however, it can be enhanced if dark matter is dominated with sub-solar-mass primordial black holes~\cite{cusin}. Furthermore, the directionality dependence of the SGWB have been recently explored~\cite{bartolo,pitrou}.

The method adopted in current GW experiments for detecting SGWB is to correlate the responses of a pair of detectors to the GW strain amplitude. This allows us to filter out detector noises and obtain a large signal-to-noise ratio~\cite{romano}. The correlation between the GW strain data from a pair of detectors is a convolution of the sky map of the SGWB with the overlap reduction function (ORF)~\cite{michelson1987,christensen1992,flanagan1993,allen1997,allen1999,cornish2001,seto2006,seto2007,seto2008,thrane2009,crowder2013,romano2017}. By correlating outputs from two different GW detectors, it is possible to detect these intensity and polarization anisotropies or the Stokes parameters of the SGWB. In this article, we provide an unified framework to calculate the ORFs for the Stokes parameters in the spherical harmonic basis.

\section{Formalism}

In the Minkowskian vacuum, the metric perturbation $h_{ij}$ in the transverse traceless gauge depicts GWs propagating at the speed of light $c=\omega/k$. At a given spacetime point $(t,\vec{x})$, it can be expanded in terms of its Fourier modes:
\be
\label{eq:planwave}
h_{ij}(t,\vec{x}) = \sum_{A}\intinf \d f \int_{S^2} \d\hat{k} \;
    h_A(f,\hat{k}) \mathbf{e}^A_{ij}(\hat{k})
    e^{-2 \pi i f (t - \hat{k}\cdot \vec{x}/c)}\,,
\ee
where $A$ stands for the polarization or the helicity of GWs described by the corresponding basis tensors $\mathbf{e}^A_{ij}(\hat{k})$, which are transverse to the direction of the wave propagation denoted by $\hat{k}$. Since $h_{ij}$ is real, its Fourier components are not fully independent with each other. For our application, we require those Fourier components with negative frequencies to be $h_A(-f,\hat{k})=h_A^*(f,\hat{k})$ for all $f\ge0$. The GWs are considered as stochastic as long as $h_{ij}$ are random fields thus characterized by their ensemble averages. Besides, assuming the probability distribution of the random amplitude $h_{ij}$ be Gaussian, then only the two-point correlation function $\langle h_{ij}(t,\vec{x}_1) h_{ij}(t,\vec{x}_2) \rangle$ is needed to describe its statistical behavior. Furthermore, if the waves are homogeneous, i.e., having translational symmetry, the ensemble average can be evaluated by doing spatial averages. As a result, the two-point correlation function of the Fourier modes should have the following form  
\be
\label{eq:paa}
\langle h_{A}(f,\hat{k}) h^*_{A'}(f',\hat{k}') \rangle
= \delta(f-f') \delta(\hat{k}-\hat{k}')P_{AA'}(f,\hat{k}) \,,
\ee
where the $\delta(f-f')$ arises from the delta function of the magnitude of their 3-momenta $\delta(\vec{k}-\vec{k}')$ and the assumption made in Eq.~(\ref{eq:planwave}) that these waves satisfy the equation of motion in vacuum. Also, the presence of $\delta(f-f')$ implies that the signal is stationary. This is a fairly good approximation during an observing period for a typical experiment. For example, a period of about 9 months of the LIGO second observing run (O2) contains 99 days of clean data for searching the isotropic background~\cite{ligo2019_O2iso} and the directional search~\cite{ligo2019_O2aniso}.

For GWs coming from the sky direction $-\hat{k}$ with wave vector $\vec{k}$, it is customary to write the polarization basis tensors in terms of the basis vectors in the spherical coordinates:
\begin{align}
    \mathbf{e}^{+}(\hat{k}) &= \hat{\mathbf{e}}_\theta \otimes \hat{\mathbf{e}}_\theta  
                               - \hat{\mathbf{e}}_\phi \otimes \hat{\mathbf{e}}_\phi \,, \\
    \mathbf{e}^{\times}(\hat{k}) &= \hat{\mathbf{e}}_\theta \otimes \hat{\mathbf{e}}_\phi  
                                    + \hat{\mathbf{e}}_\phi \otimes \hat{\mathbf{e}}_\theta  \,,
\end{align}
in which $\hat{\mathbf{e}}_\theta$, $\hat{\mathbf{e}}_\phi$, and $\hat{k}$ form a right-handed orthonormal basis.
Also, we can define the complex circular polarization basis tensors as
\begin{align}
    \mathbf{e}_{R} &= \frac{(\mathbf{e}_{+} + i \mathbf{e}_{\times})}{\sqrt{2}} \,,
   &\mathbf{e}_{L} &= \frac{(\mathbf{e}_{+} - i \mathbf{e}_{\times})}{\sqrt{2}} \,,
\end{align}
where $\mathbf{e}_{R}$ stands for the right-handed GW with a positive helicity while $\mathbf{e}_{L}$ stands for the left-handed GW with a negative helicity. The corresponding amplitudes in Eq.~(\ref{eq:planwave}) in the two different bases are related to each other via:
\begin{align}
    h_{R} &= \frac{(h_{+} - i h_{\times})}{\sqrt{2}} \,,
   &h_{L} &= \frac{(h_{+} + i h_{\times})}{\sqrt{2}} \,.
\end{align}

Analogous to the case in electromagnetic waves~\cite{book:BornAndWolf}, the coherency matrix $P_{AA'}$ in Eq.~(\ref{eq:paa}) is related to the Stokes parameters, $I$, $Q$, $U$, and $V$ as 
\begin{align}
    I &= \left[ \langle h_R h_R^* \rangle + \langle h_L h_L^* \rangle \right] / 2 \,,\\
    Q + iU &=  \langle h_L h_R^*  \rangle \,,\\
    Q - iU &=  \langle h_R h_L^*  \rangle \,,\\
    V &= \left[ \langle h_R h_R^* \rangle - \langle h_L h_L^* \rangle \right] / 2 \,.
\end{align}
They are functions of the frequency, $f$, and the propagation direction, $\hat{k}$.
To get some flavor of the meaning of these Stokes parameters, we may take a look at an example for an unpolarized quasi monochromatic GW signal with a constant intensity. It should have a constant $I$ which represents the total intensity regardless of its polarization. We have $V=0$ since the power in the right-handed and the left-handed modes should be identical. Also, because the relative phase between the left-handed and the right-handed modes ($\arg(h_L)-\arg(h_R)$) is random for an unpolarized source, the ensemble average $\langle h_L h_R^* \rangle \sim \langle e^{i(\arg(h_L)-\arg(h_R))} \rangle$ becomes zero, thereby making $Q=U=0$. 

Presumably, if one can point a GW telescope with a finite resolution and a polarization capability to a certain direction on the sky, it would be possible to measure the Stokes parameters of the incoming GWs from different patches of the sky. Unfortunately, neither a physical GW polarizer nor a directional GW detector is feasible with current technology. Alternatively, we may extract these anisotropies by combining or correlating the outputs from different existing GW detectors.

GW detectors that use laser interferometers to measure the differential length change along two different directions, such as LIGO, Virgo, and KAGRA, provide the so-called strain data $s(t) = h(t) + n(t)$ representing the fractional change of the differential arm length, where $h(t)$ is the signal due to the GW and $n(t)$ considered as noise is anything else than the signal.
The signal $h_a(t_a,\vec{x}_a)$ in a GW detector $a$ located at $\vec{x}_a$ can be expressed as the contraction of the metric perturbation $h_{ij}(t,\vec{x})$ and the detector tensor $d_a^{ij}$ of the detector:
\bw
\begin{align}
    h_a(t_a,\vec{x}_a) 
    &= d_a^{ij} h_{ij}(t_a,\vec{x}_a) \nonumber \\
    &= 
        d_a^{ij}
        \sum_{A}\intinf \d f \int_{S^2} \d\hat{k} \;
        h_A(f,\hat{k}) \mathbf{e}^A_{ij}(\hat{k})
        e^{-2 \pi i f (t_a - \hat{k}\cdot \vec{x}_a/c)} \,,
\end{align}
\ew
where the detector tensor is
\begin{align}
    d_a^{ij} &= 
    \frac{1}{2}
    \left(
    \mathbf{X}_a^{i}
    \mathbf{X}_a^{j}-
    \mathbf{Y}_a^{i}
    \mathbf{Y}_a^{j}
    \right) \,,
\end{align}
with $\mathbf{X}_a^{i}$ being the $i$-th component of the unit vector along the X-arm of the detector, while $\mathbf{Y}_a^{i}$ representing the Y-arm.

\begin{figure}
\centering
\includegraphics[width=0.5\textwidth]{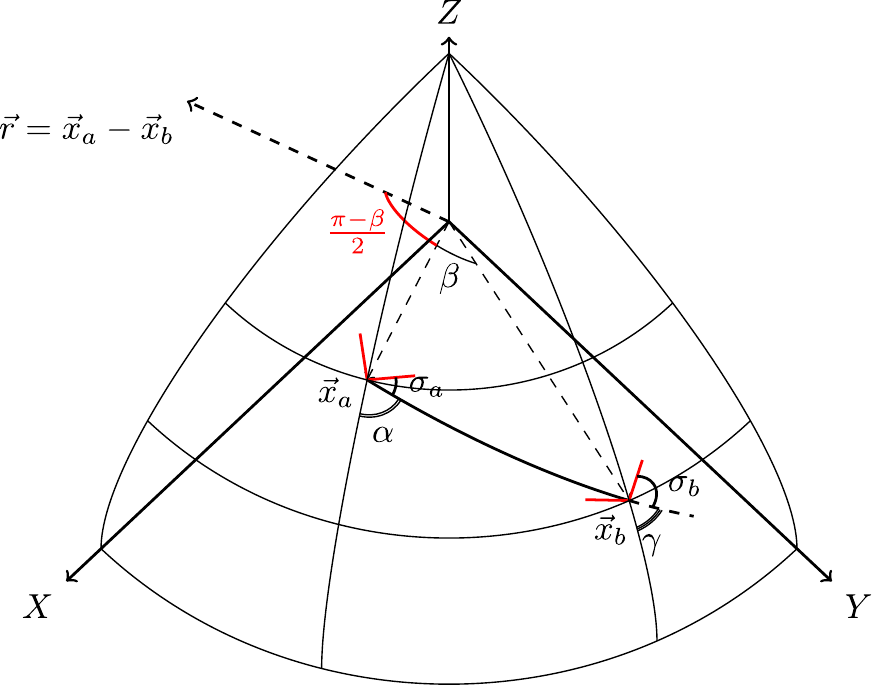}
    \caption[Coordinate]{Convention of Angles. $\vec{x}_a$, $\vec{x}_b$ represent the positions of detector-a and detector-b, respectively. $\vec{r}$ is the baseline. $\sigma_a$ and $\sigma_b$ are the angles between the great circle connecting the pair a-b and the X-arms of detector-a and detector-b, respectively.}
\label{fig:angles}
\end{figure}

The correlation of signals in a pair of detectors $a$ and $b$ can be expressed in terms of the baseline vector $\vec{r}\equiv\vec{x}_a - \vec{x}_b$ and the time delay $\tau \equiv t_a-t_b$. In frequency domain, one have
\bw
\begin{align}
   \xi_{ab}( f, \vec{r} ) 
  =& 
    \int_{-T/2}^{T/2} \d \tau \;
    \langle h_a(t_a,\vec{x}_a) h_b^*(t_b,\vec{x}_b) \rangle
    \; e^{2\pi i f \tau} 
    \nonumber \\
  =&
    d_a^{ij} d_b^{kl}
    \int_{S^2} \d \hat{k} \;
    \sum_{AA'}
    P_{AA'}(f,\hat{k})
    \mathbf{e}^{A}_{ij}(\hat{k}) \mathbf{e}^{*A'}_{kl}(\hat{k})
    e^{2\pi i f (\hat{k}\cdot\vec{r}/c)} 
   \nonumber \\
  =&
    \label{eq:xi2gamma}
    \sum_{S=\{I,V,Q\pm iU\}}
    \int_{S^2} \d \hat{k} \;
    S(f,\hat{k}) \;
    {}^{ijkl}\mathbb{D}_{ab} \;
    \mathbb{E}^{S}_{ijkl} (\hat{k}) 
    e^{2\pi i f (\hat{k}\cdot\vec{r}/c)} \,,
\end{align}
\ew
where the Fourier integral is taken over an interval $T$ within which the orientation and the condition of the detectors are approximately fixed. 
In addition, the interval $T$ has to be large enough when compared with the period of GW signals in the detectors.
In Eq.~(\ref{eq:xi2gamma}), the polarization tensors $\mathbb{E}$ associated with the corresponding Stokes parameters are defined as
\begin{align}
    \label{eq:EEII}
    \mathbb{E}^{I}_{ijkl} (\hat{k}) &= 
    \mathbf{e}^{R}_{ij}(\hat{k}) \mathbf{e}^{*R}_{kl}(\hat{k}) + 
    \mathbf{e}^{L}_{ij}(\hat{k}) \mathbf{e}^{*L}_{kl}(\hat{k}) \,,
    \\ 
    \mathbb{E}^{V}_{ijkl} (\hat{k}) &= 
    \mathbf{e}^{R}_{ij}(\hat{k}) \mathbf{e}^{*R}_{kl}(\hat{k}) - 
    \mathbf{e}^{L}_{ij}(\hat{k}) \mathbf{e}^{*L}_{kl}(\hat{k}) \,,
    \\ 
    \mathbb{E}^{Q+iU}_{ijkl} (\hat{k}) &= 
    \mathbf{e}^{L}_{ij}(\hat{k}) \mathbf{e}^{*R}_{kl}(\hat{k}) \,,
    \\ 
    \label{eq:EEQ-iU}
    \mathbb{E}^{Q-iU}_{ijkl} (\hat{k}) &= 
    \mathbf{e}^{R}_{ij}(\hat{k}) \mathbf{e}^{*L}_{kl}(\hat{k}) \,,
\end{align}
while $\mathbb{D}$ denotes the direct product of two detector tensors
\begin{align}
    \mathbb{D}(\mathbf{R}_{\mathbb{D}};\mathbf{R}_{ab})
    =
    {}^{ijkl}\mathbb{D}_{ab}
    &\equiv 
    d_a^{ij} d_b^{kl} \,,
\end{align}
which is a function of the orientations of the two detectors relative to the sky, determined by two three-dimensional rotations $\mathbf{R}_{\mathbb{D}}$ and $\mathbf{R}_{ab}$. 
To be more specific, for a pair of interferometry detectors, the fundamental degrees of freedom regarding to its geometry include the opening angle between the two arms of each detector, the orientation of each detector relative to the sky, and the baseline vector connecting the two detectors. In practice, the opening angles are usually fixed. It is $90^{\circ}$
for LIGO, Virgo, KAGRA, and Cosmic Explorer, and $60^{\circ}$ for LISA, and Einstein Telescope. As the relative orientation between the two detectors and the baseline vector are fixed, it is convenient to factor out an overall SO(3) rotation of the whole pair, which can be realized by three Euler angles. In the literature, we use $\mathbf{R}_{\mathbb{D}}$, which is an element of SO(3), to represent such an overall rotation.   
Furthermore, for a pair of ground-based detectors, we can choose the polar coordinates of the first detector $(\theta_a, \phi_a)$ and the angle $\alpha$, which gives the direction pointing to the second detector, as the rotational angles of $\mathbf{R}_{\mathbb{D}}(\phi_a, \theta_a, \alpha)$. The rest degrees of freedom can be described by the other three angles, $\sigma_a$, $\sigma_b$, and $\beta$, as illustrated in Fig.~\ref{fig:angles}. The numerical values of the six angles for detector pairs among LIGO, Virgo, and KAGRA are listed in Table~\ref{table:angles}.  In this way, the baseline vector $\vec{r}$ can be determined by $\mathbf{R}_{\mathbb{D}}$ and $\beta$. To simplify the expression, we use $\mathbf{R}_{ab}$ to denote those internal angles, i.e.~$\sigma_a$, $\sigma_b$, and $\beta$.
With the help of these angular parameters, we have
\begin{align}
    {}^{ijkl}\mathbb{D}_{ab} 
    &= d^{ij}_a(\theta_a, \phi_a, \sigma_a) 
    d^{kl}_b(\theta_b, \phi_b, \sigma_b)
   \nonumber \\
    &= 
    {\mathbf{R}_\mathbb{D}}^{ijkl}_{pqrs}
    \;
    {}^{pqrs}\mathbb{D}_{0}(\mathbf{R}_{ab}) \,,
\end{align}
where
\begin{align}
    \label{eq:D0}
    \mathbb{D}_{0}(\mathbf{R}_{ab})
    &\equiv 
    \left[
    \mathbf{R}_Z(\sigma_a)
    d_{0}
    \right]
    \otimes
    \left[
    \mathbf{R}_Y(\beta)
    \mathbf{R}_Z(\sigma_b)
    d_{0}
    \right]
\end{align}
denotes the direct product of the detector tensor pair a-b when we rotate the pair of detectors such that the detector-a is located at the north pole of the Earth while the detector-b is stayed on the $\phi=0$ meridian. In this configuration, the corresponding baseline direction $\hat{r}_0$ is  
\be
\label{eq:r0}
\hat{r}_0 = (\theta_{r_0},\phi_{r_0})=(\frac{\beta-\pi}{2}, 0)=(\frac{\pi-\beta}{2}, \pi)\,.
\ee

In the expression, the
$d_0=\frac{1}{2}(\hat{X}\otimes\hat{X}-\hat{Y}\otimes\hat{Y})$
is the detector tensor for a detector located at the north pole with its X-arm pointing to the X-axis of the celestial coordinate system, 
while the $\mathbf{R}(\alpha,\beta,\gamma) \equiv \mathbf{R}_Z(\alpha) \mathbf{R}_Y(\beta) \mathbf{R}_Z(\gamma)$ is the Euler rotation matrix, and $\mathbf{R}_X$, $\mathbf{R}_Y$, $\mathbf{R}_Z$ are three-dimensional rotation matrices that actively rotate tensors around fixed celestial $X$, $Y$, and $Z$-axes correspondingly. A brief review of the Euler rotation is given in the Appendix~\ref{sec:3drot}.
\begin{table}
    \begin{tabular}{@{\quad}c@{\quad}r@{\quad}r@{\quad}r@{\quad}r@{\quad}r@{\quad}r@{\quad}}
    \hline
    \hline
        Detectors  &$\theta_a$  &$\phi_a(t=0)$  &$\alpha$  &$\beta$   &$\sigma_a$  &$\sigma_b$  \\
    \hline                                                       
        K-H        &53.6        &137.3     &135.3     &72.4      &-15.7       &160.7       \\
        K-L        &53.6        &137.3     &139.5     &99.3      &-19.9       &250.4       \\
        V-K        &46.4        &10.5      &139.8     &86.5      &20.8        &84.1        \\
        L-V        &59.4        &-90.8     &133.2     &76.8      &154.5       &100.4       \\
        H-L        &43.5        &-119.4    &64.4      &27.2      &151.6       &241.5       \\
        H-V        &43.5        &-119.4    &145.6     &79.6      &70.4        &128.1       \\
    \hline
    \hline
    \end{tabular}
    \caption{Angular parameters in degrees for different pairs of detectors formed by KAGRA(K), Virgo(V), LIGO-Hanford(H), and LIGO-Livingston(L). The data is converted from {\bf LALSuite}~\cite{code:lalsuite} assuming that the Earth is a perfect sphere.}
    \label{table:angles}
\end{table}

To investigate, for given $I(f,\hat{k})$, $Q(f,\hat{k})$, $U(f,\hat{k})$, and $V(f,\hat{k})$, how $\xi_{ab}$ vary with the geometrical configuration or the orientation of the detector pair a-b, it is convenient to rewrite the convolutional integral Eq.~(\ref{eq:xi2gamma}) into the following form: 
\bw
\begin{align}
    \label{eq:xiSO3}
    \xi_{ab}(f,\mathbf{R}_{\mathbb{D}};\mathbf{R}_{ab})
    = 
    \sum_{S=\{I,V,Q\pm iU\}}
    \int_{S^2} \d \hat{k} \; S(f,\hat{k}) \gamma^{S}(\hat{k},f,\mathbf{R}_{\mathbb{D}};\mathbf{R}_{ab}) \,,
\end{align}
\ew
where 
\begin{align}
    \label{eq:gamma}
    \gamma^{I,V,Q\pm iU}_{ab}(\hat{k},f)
    = 
    \mathbb{D}(\mathbf{R}_{\mathbb{D}};\mathbf{R}_{ab})
    \cdot
    \mathbb{E}^{I,V,Q\pm iU}(\hat{k}) 
    e^{2\pi i f (\hat{k}\cdot\vec{r}/c)} 
\end{align}
are the kernels that convert the SGWB distribution over the whole sky into the correlation $\xi$, and are usually called the ORFs. The $ \mathbb{D} \cdot \mathbb{E}$ gives the projection of the metric perturbation into the length perturbation of each detector, while $e^{2\pi i f (\hat{k}\cdot\vec{r}/c)}$ is the phase delay of GW signals between two detectors caused by the GW traveling time.

In many cases, it is convenient to evaluate the integral in Eq.~(\ref{eq:xiSO3}) in the spherical harmonic basis:
\begin{align}
   \label{eq:xilm} 
   \xi_{ab}(f,\mathbf{R}_{\mathbb{D}};\mathbf{R}_{ab})
   &=
    \sum_{S=\{I,V,Q\pm iU\}}
   \sum_{\ell m}  S_{\ell m}(f) \gamma_{\ell m}^{S}(f,\mathbf{R}_{\mathbb{D}};\mathbf{R}_{ab}) \,,
\end{align}
where we have expanded the Stokes parameters in terms of ordinary and spin-weighted spherical harmonics as
\begin{align}
    \label{eq:IIlm}
    I(f,\hat{k}) &= \sum_{\ell m}I_{\ell m}(f) \; Y_{\ell m}(\hat{k}) \,,\\
    \label{eq:VVlm}
    V(f,\hat{k}) &= \sum_{\ell m}V_{\ell m}(f) \; Y_{\ell m}(\hat{k}) \,,\\
    \label{eq:Q+iUlm}
    (Q+iU)(f,\hat{k}) &= \sum_{\ell m}(Q+iU)_{\ell m}(f) \; _{+4}Y_{\ell m}(\hat{k}) \,,\\
    \label{eq:Q-iUlm}
    (Q-iU)(f,\hat{k}) &= \sum_{\ell m}(Q-iU)_{\ell m}(f) \; _{-4}Y_{\ell m}(\hat{k}) \,,
\end{align}
so as the ORFs:
\begin{align}
    \label{eq:gammaIVlm}
    \gamma_{\ell m}^{I,V}(f,\mathbf{R}_{\mathbb{D}};\mathbf{R}_{ab}) 
    &=\int_{S^2} \d \hat{k} \; Y_{\ell m}(\hat{k}) \gamma^{I,V}(\hat{k},f,\mathbf{R}_{\mathbb{D}};\mathbf{R}_{ab}) \,,\\
    \label{eq:gammaQUlm}
    \gamma_{\ell m}^{Q\pm i U}(f,\mathbf{R}_{\mathbb{D}};\mathbf{R}_{ab}) 
    &=\int_{S^2} \d \hat{k} \; _{\pm4}Y_{\ell m}(\hat{k}) \gamma^{Q\pm iU}(\hat{k},f,\mathbf{R}_{\mathbb{D}};\mathbf{R}_{ab})\,.
\end{align}
The specific combinations, $Q\pm iU$, make them become spin $\pm4$ objects so that we can expand them nicely by the corresponding spin-weighted spherical harmonics. 

By plugging Eq.~(\ref{eq:gamma}) into Eqs.~(\ref{eq:gammaIVlm}) and~(\ref{eq:gammaQUlm}), and expanding the polarization basis tensors as
\begin{align}
    \label{eq:EEIIlm}
    \mathbb{E}^{I}_{ijkl} (\hat{k}) 
    &=\sum_{\ell_e m_e} {}_{ijkl}\mathbb{E}^{I}_{\ell_e m_e} Y_{\ell_e m_e}(\hat{k}) \,,
   \\
    \mathbb{E}^{V}_{ijkl} (\hat{k}) 
    &=\sum_{\ell_e m_e} {}_{ijkl}\mathbb{E}^{V}_{\ell_e m_e} Y_{\ell_e m_e}(\hat{k}) \,,
   \\
    \mathbb{E}^{Q+iU}_{ijkl} (\hat{k}) 
    &=\sum_{\ell_e m_e} {}_{ijkl}\mathbb{E}^{Q+iU}_{\ell_e m_e} {}_{-4}Y_{\ell_e m_e}(\hat{k}) \,,
   \\
    \label{eq:EEQ-iUlm}
    \mathbb{E}^{Q-iU}_{ijkl} (\hat{k}) 
    &=\sum_{\ell_e m_e} {}_{ijkl}\mathbb{E}^{Q-iU}_{\ell_e m_e} {}_{+4}Y_{\ell_e m_e}(\hat{k}) \,,
\end{align}
we can express $\xi$ in the following form, in which $s=0,\pm4$ correspond to their respective Stokes parameters:
\bw
\begin{align}
    \gamma_{\ell m}(f,\mathbf{R}_{\mathbb{D}};\mathbf{R}_{ab})
  &=
    \int \d \hat{k} \;
    _{-s}Y_{\ell m}(\hat{k}) \;
    \mathbb{D}(\mathbf{R}_{\mathbb{D}};\mathbf{R}_{ab})
    \cdot
    \mathbb{E}(\hat{k}) \;
    e^{2\pi i f (\hat{k}\cdot\vec{r}/c)} 
    \nonumber \\
  &=
    {}^{ijkl}\mathbb{D}(\mathbf{R}_{\mathbb{D}};\mathbf{R}_{ab})
    \int \d \hat{k} \;
    {}_{-s}Y_{\ell m}(\hat{k}) \;
    \sum_{\ell_e m_e}
    {}_{ijkl}\mathbb{E}_{\ell_e m_e} \;
    {}_{s}Y_{\ell_e m_e}(\hat{k}) \;
    (4\pi) \sum_{LM} i^L j_L(\frac{2\pi f r}{c}) Y_{LM}^*(\hat{k}) Y_{LM}(\hat{r})
    \nonumber \\
  &=
    \label{eq:gamma_old}
    {}^{ijkl}\mathbb{D}(\mathbf{R}_{\mathbb{D}};\mathbf{R}_{ab})
    \sum_{\ell_e m_e}
    {}_{ijkl}\mathbb{E}_{\ell_e m_e} \;
    (4\pi) \sum_{LM} i^L j_L(\frac{2\pi f r}{c}) 
    Y_{LM}(\hat{r})
    \left\langle
    \begin{matrix}
        L  && \ell_e   && \ell \\
        M  && s\; m_e  && -s\;  m
    \end{matrix}
    \right\rangle \,,
\end{align}
where we have used the shorthand notation for the integral of three spherical harmonics given by
\begin{align}
    \left\langle
    \begin{matrix}
        L && l_1  &&  l_2 \\
        M && s_1\;m_1  && s_2\;m_2 
    \end{matrix}
    \right\rangle
    &\equiv
    \int \d \hat{k} \;
    Y_{LM}^*(\hat{k})\; {}_{s_1}\!Y_{l_1 m_1}(\hat{k}) \; {}_{s_2}\!Y_{l_2 m_2}(\hat{k})
    \nonumber \\
    &=
    \label{eq:threeJ}
    (-1)^M
    \sqrt{\frac{(2L+1)(2l_1+1)(2l_2+1)}{4\pi}}
    \begin{pmatrix}
        L && l_1  &&  l_2 \\
        0 && -s_1  &&  -s_2 
    \end{pmatrix}
    \begin{pmatrix}
        L && l_1  &&  l_2 \\
       -M && m_1  &&  m_2 
    \end{pmatrix} \,,
\end{align}
\ew
which involves two Wigner-3j symbols representing the coupling coefficients between different spherical harmonics~\cite{book:Varshalovich}.  
It is worth noting that the properties of the Wigner-3j symbols in Eq.~(\ref{eq:threeJ}) imply that $L$, $\ell_e$, and $\ell$ have to satisfy the triangular condition, i.e.~$\ell + \ell_e \ge L \ge \ell - \ell_e$, while $-M+m_e+m=0$. 

Nevertheless, a rotation of the pair of detectors on the Earth is equivalent to rotating the sky in the reverse sense. This fact enables us to choose a convenient coordinate system to evaluate Eq.~(\ref{eq:gamma_old}). A convenient choice is to place the pair of detectors in the position described by Eq.(\ref{eq:D0}). Equivalently, we can perform the rotation $\mathbf{R}^{-1}_{\mathbb{D}}$ on the detector pair, the baseline, and the SGWB sky simultaneously by using Eqs.~(\ref{eq:rotYlm}) and~(\ref{eq:rotsYlm}), turning $\hat{k}$ and $\hat{r}$ into $\mathbf{R}^{-1}_{\mathbb{D}} \hat{k}$ and $\mathbf{R}^{-1}_{\mathbb{D}} \hat{r}=\hat{r}_0$, respectively. In this coordinate system, the explicit forms of $\gamma^I_{\ell m}$, $\gamma^{Q+iU}_{\ell m}$, $\gamma^{Q-iU}_{\ell m}$, and $\gamma^V_{\ell m}$ are given by
\bw
\begin{align}
    \label{eq:gammaIV_lm}
    \gamma_{\ell m}^{I,V}(f,\mathbf{R}_{\mathbb{D}};\mathbf{R}_{ab}) 
  &=
    (4\pi) 
    \sum_{m'}
    D^{\ell}_{m' m}(\mathbf{R}^{-1}_{\mathbb{D}}) 
    \sum_{\ell_e m_e}
    \mathbb{D}_0(\mathbf{R}_{ab})
    \cdot
    \mathbb{E}^{I,V}_{\ell_e m_e} 
    \sum_{L M} 
    i^L j_L(\frac{2\pi f r}{c}) 
    Y_{L M}(\hat{r}_0)  
    \left\langle
    \begin{matrix}
        L  && \ell_e   && \ell \\
        M  && 0\; m_e  && 0\;  m'
    \end{matrix}
    \right\rangle 
    \,, \\
    \label{eq:gammaQU_lm}
    \gamma_{\ell m}^{Q\pm iU}(f,\mathbf{R}_{\mathbb{D}};\mathbf{R}_{ab}) 
  &=
    (4\pi) 
    \sum_{m'}
    D^{\ell}_{m' m}(\mathbf{R}^{-1}_{\mathbb{D}}) 
    \sum_{\ell_e m_e}
    \mathbb{D}_0(\mathbf{R}_{ab})
    \cdot
    \mathbb{E}^{Q\pm iU}_{\ell_e m_e} 
    \sum_{L M} 
    i^L j_L(\frac{2\pi f r}{c}) 
    Y_{L M}(\hat{r}_0)  
    \left\langle
    \begin{matrix}
        L  && \ell_e        && \ell \\
        M  && \mp4\; m_e    && \pm4\;  m'
    \end{matrix}
    \right\rangle \,,
\end{align}
\ew
respectively, where $D^{\ell}_{m' m}(\mathbf{R}^{-1}_{\mathbb{D}})=D^{\ell}_{m' m}(-\alpha,-\theta_a,-\phi_a)$.
Using Eqs.~(\ref{Yconjugate}),~(\ref{Yparity}), and the conjugate relations for the antenna pattern functions in Appendix~\ref{sec:DE}, it is straightforward to show that the four ORFs have the conjugate relations:
\begin{align}
\gamma_{\ell -m}^{I,V}=&(-1)^{\ell+m}\gamma_{\ell m}^{I,V*}\,, \label{IVconjugate}\\
\gamma_{\ell -m}^{Q\pm iU}=&(-1)^{\ell+m}\gamma_{\ell m}^{Q\pm iU*}\,. \label{QUconjugate}
\end{align}
In Eqs.~(\ref{eq:gammaIV_lm}) and~(\ref{eq:gammaQU_lm}), the Wigner-D matrices $D^{\ell}_{m' m}(\mathbf{R}^{-1}_{\mathbb{D}})$ account for the degrees of freedom reflecting the free rotation of the whole pair of detectors. In the case of ground-based GW detectors, the two Euler angles, $\theta_a$ and $\alpha$, are fixed with respect to the geographical locations of the detectors, while $\phi_a$ changes azimuthally as the Earth rotates. Besides, the frequency dependency of these $\gamma$'s, caused by the time delay of GW signal arriving at each detector, is taken cared of by the projection into the spherical Bessel functions $j_L(2 \pi f r/c)$.  

\bw

\section{Isotropic Overlap Reduction Functions}

\subsection{Unpolarized Isotropic Case}

In the case of isotropic and unpolarized SGWB, the only relevant ORF is the $\gamma_{00}^{I}$, which can be calculated from Eq.~(\ref{eq:gammaIV_lm}) as
\begin{align}
    \gamma_{00}^{I}(f,\mathbf{R}_{\mathbb{D}};\mathbf{R}_{ab}) 
  &=
    (4\pi) 
    \sum_{m'}
    D^{0}_{m' 0}(\mathbf{R}^{-1}_{\mathbb{D}}) 
    \sum_{\ell_e m_e}
    \mathbb{D}_0(\mathbf{R}_{ab})
    \cdot
    \mathbb{E}^{I}_{\ell_e m_e} 
    \sum_{L M} 
    i^L j_L(\frac{2\pi f r}{c}) 
    Y_{L M}(\hat{r}_0)  
    \left\langle
    \begin{matrix}
        L  && \ell_e   && 0 \\
        M  && 0\; m_e  && 0\;  m'
    \end{matrix}
    \right\rangle 
    \nonumber \\
  &=
    \sqrt{4\pi}
    \sum_{\ell_e m_e}
    \mathbb{D}_0(\mathbf{R}_{ab})
    \cdot
    \mathbb{E}^{I}_{\ell_e m_e} 
    \sum_{L M} 
    i^L j_L(\frac{2\pi f r}{c}) 
    Y_{L M}(\hat{r}_0)  
    \delta_{L \ell_e}
    \delta_{M m_e}
   \nonumber  \\
  &=
    \sqrt{4\pi}
    \sum_{\ell_e m_e}
    \mathbb{D}_0(\mathbf{R}_{ab})
    \cdot
    \mathbb{E}^{I}_{\ell_e m_e} 
    i^{\ell_e} j_{\ell_e}(\frac{2\pi f r}{c}) 
    Y_{\ell_e m_e}(\hat{r}_0)  \,.
    \label{eq:gamma_iso}
\end{align}
\ew
The result does not depend on the orientation of the whole pair, i.e., the overall SO(3) rotation $\mathbf{R}_{\mathbb{D}}$. This is due to the fact that there is no preferred direction for an isotropic unpolarized SGWB. 
In Eq.~(\ref{eq:gamma_iso}), the expression of $\mathbb{D}_0(\mathbf{R}_{ab}) \cdot \mathbb{E}^{I}_{\ell_e m_e}$ can be found in Appendix~\ref{sec:DEI}. By summing all 15 terms whose $\ell_e = 0,2,4$ and $m_e$ are integers from $-\ell_e$ to $\ell_e$, one can get the standard ORF of unpolarized isotropic SGWB as same as, e.g., the result given in Ref.~\cite{seto2008}: 
\bw
\begin{align}
    \gamma_{00}^{I}(f,\mathbf{R}_{\mathbb{D}};\mathbf{R}_{ab}) 
    =&
    \cos (2 (\sigma_1-\sigma_2))
    \nonumber
    \\
    \times&
    \frac{4\sqrt{\pi}}{5}
    \left[\left(j_0+\frac{5 j_2}{7}+\frac{3 j_4}{112}\right) \cos ^4\left(\frac{\beta }{2}\right) \right] 
    \nonumber
    \\
    +&\cos (2 (\sigma_1+\sigma_2)+\pi )
    \nonumber
    \\
    \times&
    \frac{4\sqrt{\pi}}{5}
    \left[
      \left(\frac{-3 j_0}{8} +\frac{45 j_2}{56}-\frac{169 j_4}{896}\right)
      +\left(\frac{j_0}{2}-\frac{5 j_2}{7}-\frac{27 j_4}{224}\right) \cos (\beta ) 
      +\left(-\frac{j_0}{8}-\frac{5 j_2}{56}-\frac{3 j_4}{896}\right) \cos (2 \beta )
    \right] \,.
\end{align}
\ew

\bw
\subsection{Circularly Polarized Isotropic Case}

In the case of isotropic and circularly polarized SGWB, the other relevant ORF is the $\gamma_{00}^{V}$, which can be calculated from Eq.~(\ref{eq:gammaIV_lm}) as
\be
    \gamma_{00}^{V}(f,\mathbf{R}_{\mathbb{D}};\mathbf{R}_{ab}) =
      \sqrt{4\pi} 
    \sum_{\ell_e m_e}
    \mathbb{D}_0(\mathbf{R}_{ab})
    \cdot
    \mathbb{E}^{V}_{\ell_e m_e} 
    i^{\ell_e} j_{\ell_e}(\frac{2\pi f r}{c}) 
    Y_{\ell_e m_e}(\hat{r}_0)  \,.
\ee
\ew
Similarly, using the expression of $\mathbb{D}_0(\mathbf{R}_{ab}) \cdot \mathbb{E}^{V}_{\ell_e m_e}$ in Appendix~\ref{sec:DEV}, we obtain the ORF of circularly polarized isotropic SGWB as same as the result found in Ref.~\cite{seto2008}: 
\bw
\be
    \gamma_{00}^{V}(f,\mathbf{R}_{\mathbb{D}};\mathbf{R}_{ab}) =
    -\frac{4\sqrt{\pi}}{5}  \sin (2 (\sigma_1+\sigma_2)+\pi )\, \sin\left(\frac{\beta }{2}\right) 
     \left[   \left(-j_1+\frac{7 j_3}{8}\right)
      +\left(j_1+\frac{3 j_3}{8}\right) \cos (\beta ) 
    \right] \,.
\ee
\ew
\bw

\begin{figure}[ht!]
\centering
\includegraphics[width=0.86\textwidth]{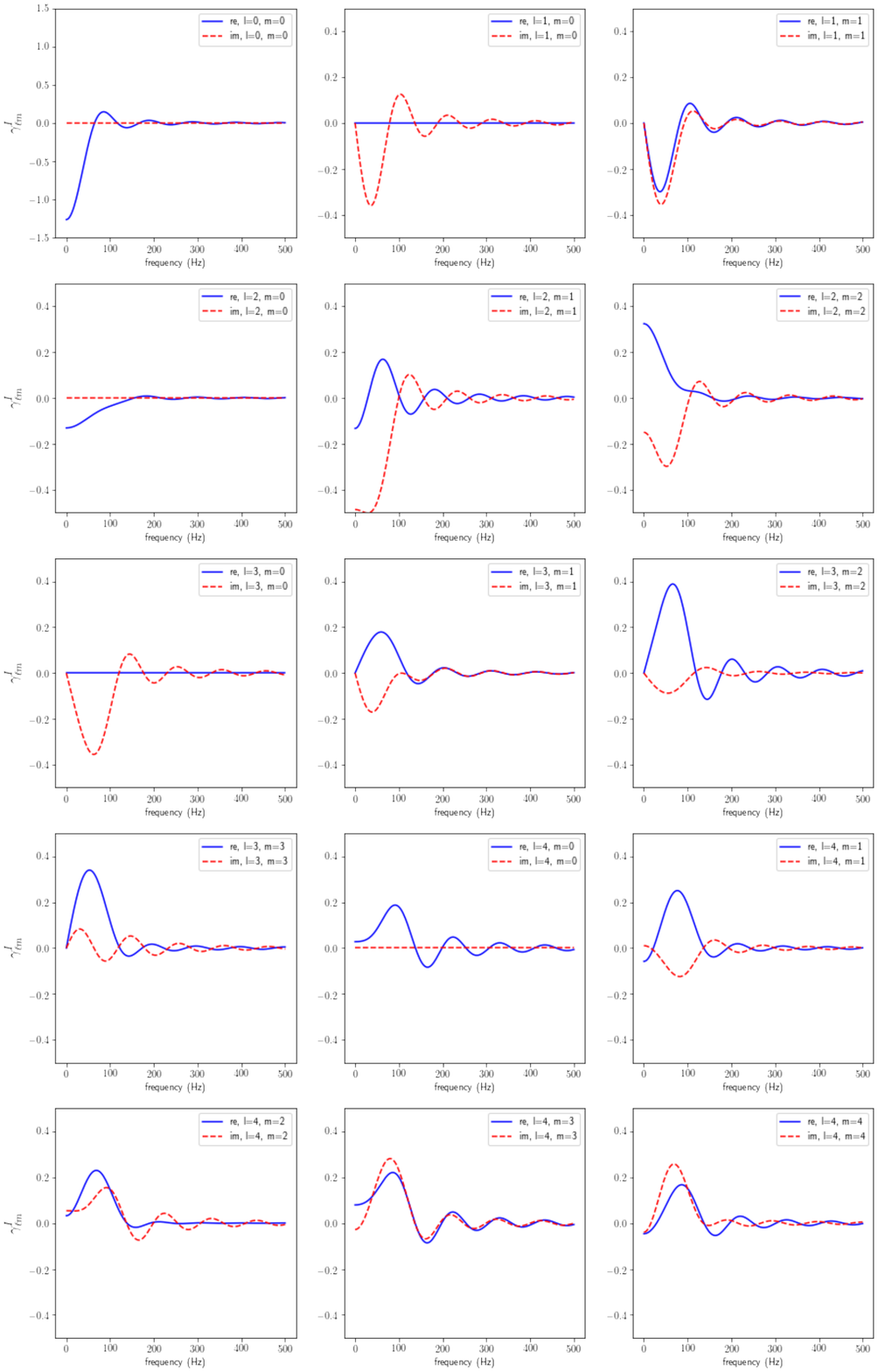}
    \caption[Coordinate]{Real and imaginary parts of the multipole moments of the intensity overlap reduction function
    $\gamma_{\ell m}^I$ for the LIGO Hanford-LIGO Livingston detector pair. 
    Plots of $\ell=0,1,2,3,4$ and $m\ge 0$ are shown. The $m<0$ multipoles can be obtained by using the conjugate relation in Eq.~(\ref{IVconjugate}).}
\label{fig:ORFI}
\index{figures}
\end{figure}

\begin{figure}[ht!]
\centering
\includegraphics[width=0.86\textwidth]{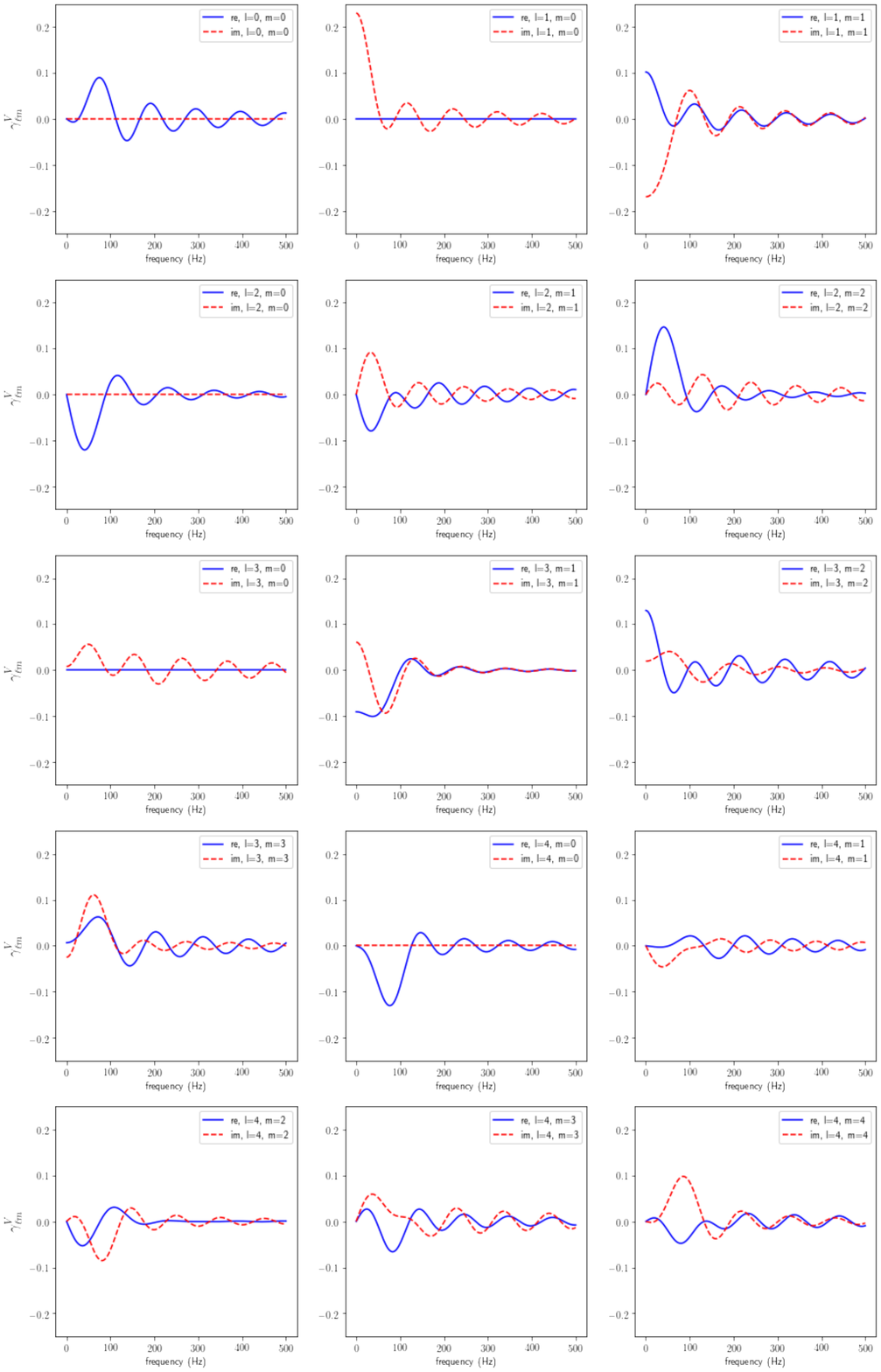}
    \caption[Coordinate]{Real and imaginary parts of the multipole moments of the circular-polarization
     overlap reduction function $\gamma_{\ell m}^V$ for the LIGO Hanford-LIGO Livingston detector pair. 
    Plots of $\ell=0,1,2,3,4$ and $m\ge 0$ are shown. The $m<0$ multipoles can be obtained by using the conjugate relation in Eq.~(\ref{IVconjugate}).}
\label{fig:ORFV}
\index{figures}
\end{figure}

\begin{figure}[ht!]
\centering
\includegraphics[width=0.78\textwidth]{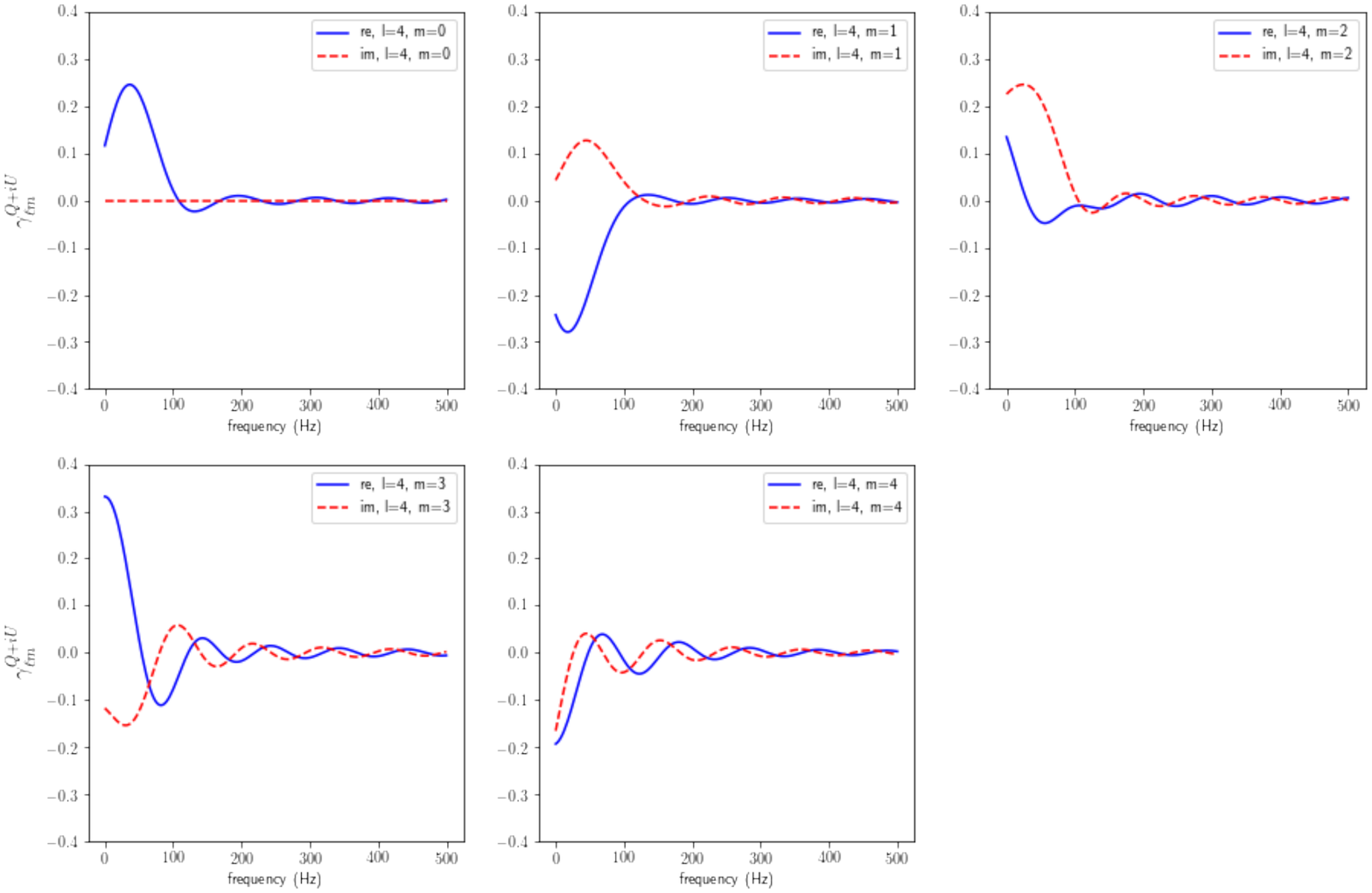}
\includegraphics[width=0.78\textwidth]{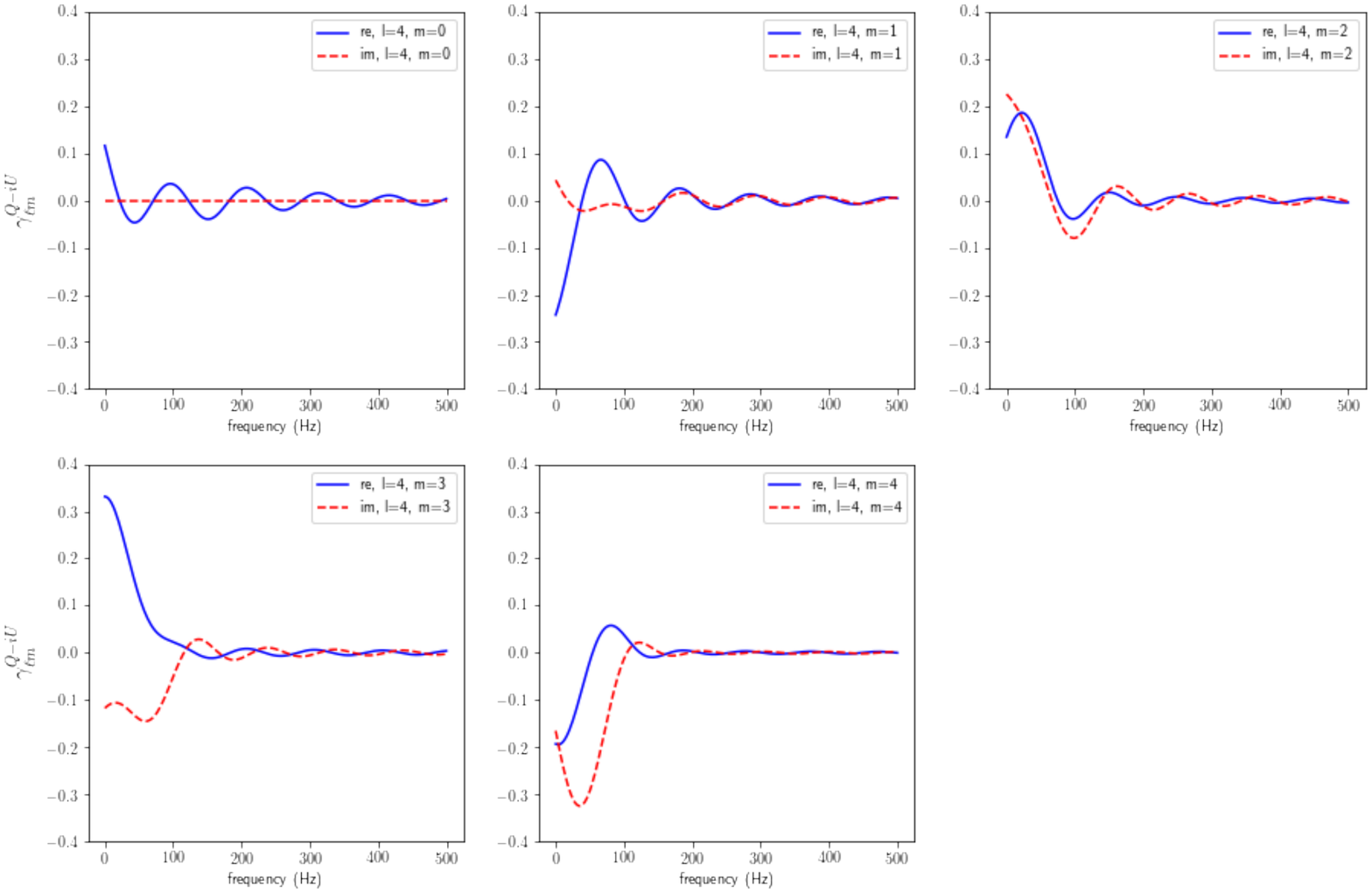}
    \caption[Coordinate]{Real and imaginary parts of the multipole moments of the linear-polarization
     overlap reduction function $\gamma_{\ell m}^{Q\pm iU}$ for the LIGO Hanford-LIGO Livingston detector pair. 
    Plots of $\ell=4$ and $m=0,1,2,3,4$ are shown. The $m<0$ multipoles can be obtained by using the conjugate relations in Eq.~(\ref{QUconjugate}).}
\label{fig:ORFQU}
\index{figures}
\end{figure}
\ew

\section{Overlap-Reduction-Function Harmonics}

We compute explicit expressions of the multipole moments of the ORFs in Eqs.~(\ref{eq:gammaIV_lm}) and~(\ref{eq:gammaQU_lm}) for the two LIGO detectors (H-L) listed in Table~\ref{table:angles},
using $\alpha=64.4^\circ$, $\theta_a=43.5^\circ$, 
and $\phi_a=-119.4^\circ$ for the Wigner-D matrices $D^{\ell}_{m' m}$,
and $\beta=27.2^\circ$, $\sigma_a=\sigma_1=151.6^\circ$, and $\sigma_b=\sigma_2=241.5^\circ$ for the 
antenna pattern functions in Appendix~\ref{sec:DE}.
The numerical results are shown in Figs.~\ref{fig:ORFI}, \ref{fig:ORFV},~and \ref{fig:ORFQU}. 
We have plotted the multipole moments for $l\le 4$, noting that higher multipoles can be easily generated by our code. 
The $\gamma_{\ell m}^{I}$'s in Fig.~\ref{fig:ORFI} and the $\gamma_{00}^{V}$ in Fig.~\ref{fig:ORFV}
match those made in Ref.~\cite{romano2017}. The ORF multipoles for circular and linear polarizations are new results of this work. In practice, the correlation output of the detector pair is a sum of contributions from all polarizations. To disentangle the contributions from the four different Stokes parameters to the observed signal would be a challenging inversion problem. The multipole expansion of the Stokes parameters and their respective ORFs may provide us with a systematic tool of tackling this inversion. The $\ell$- and frequency dependence of the ORF multipoles would
allow us to assess the sensitivity of the detector pair to different Stokes parameters at different angular scales such that efficient estimators of the anisotropy and polarization power spectra can be constructed for extracting SGWB anisotropies from observational data. Initial efforts in this direction have been put forth~\cite{seto2006,seto2008,thrane2009,romano2017}.

\begin{figure}
\centering
\includegraphics[width=0.45\textwidth]{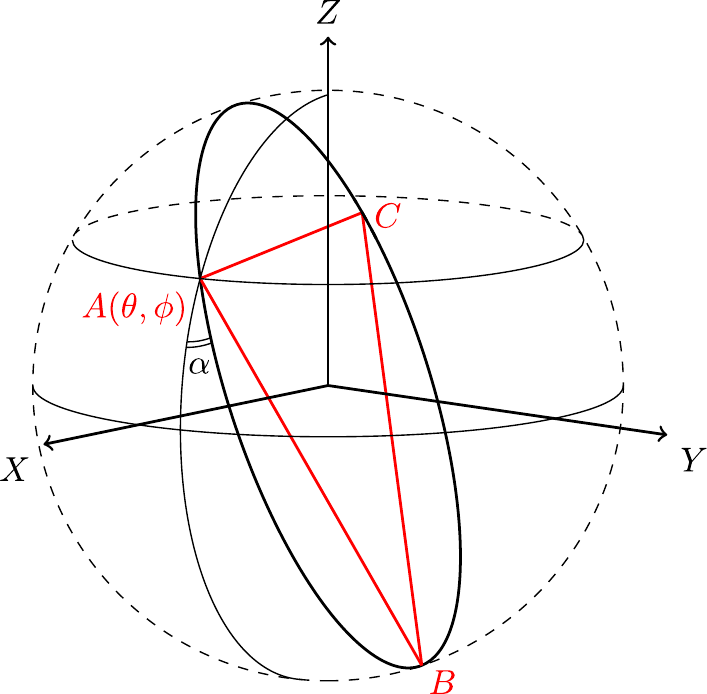}
    \caption[Coordinate]{LISA-type space-based detectors.}
\label{fig:lisa}
\end{figure}

\section{Correlation Output Data}

Ideally, a SGWB sky map can be constructed from a time series of the correlation output from a pair of GW detectors through a convolutional integral~(\ref{eq:xiSO3}) over the whole sky. Regarding to different polarizations or Stokes parameters, the corresponding kernels, i.e., the ORFs, can be expanded in terms of the spherical harmonic basis to convert the convolutional integral into a summation over multipole moments in Eq.~(\ref{eq:xilm}). The multipole moments of the ORFs are given in Eqs.~(\ref{eq:gammaIV_lm}) and~(\ref{eq:gammaQU_lm}). They depend on an overall SO(3) rotation $\mathbf{R}_{\mathbb{D}}$ characterizing the relative orientation between the whole pair of detectors and the SGWB sky, the antenna pattern function $\mathbb{D} \cdot \mathbb{E}$ in the unrotated frame defined by Eqs.~(\ref{eq:D0}) and~(\ref{eq:r0}), and the spherical Bessel function $j_L(2\pi f r/c)$ term that gives the frequency dependency of the ORF for a given baseline vector. All these three terms are coupled through the coefficients~(\ref{eq:threeJ}) involving Wigner-3j symbols.
Thus, each pointing on the SGWB sky map has a time-accumulated data output in an element of three Euler angles $(\alpha,\theta_a,\phi_a)$ in the group manifold of the three-dimensional rotation. Indeed, this data structure is similar to that in the observation of the temperature anisotropy and polarization of the cosmic microwave background with an asymmetric beam pattern~\cite{challinor2000,wandelt2001}. We can borrow the fast algorithms developed from there for simulation of interferometry experiments as well as analysis of the experimental data. Here we will give a brief outline and leave the details in the future work. The key to the algorithms is to factor the Wigner D-matrix:
\be
D(\alpha,\theta_a,\phi_a)=D(\phi_a-\pi/2,-\pi/2,\theta_a)\,D(0,\pi/2,\alpha+\pi/2).
\ee
As such, the Euler angles $(\alpha,\theta_a,\phi_a)$ only appear in complex exponentials, so the full three-sphere of rotations can now be calculated with a three-dimensional fast Fourier transform. The Fourier components of the correlation output are then given by
\bw
\be
T_{mm'm''}=\frac{1}{(2\pi)^3}\int_0^{2\pi} d\alpha\,d\theta_a\,d\phi_a\,\xi_{ab}(\alpha,\theta_a,\phi_a)\,e^{-im\alpha-im'\theta_a-im''\phi_a},
\ee
\ew
which can be used for extracting the angular power spectra of each of the four Stokes parameters. 

In reality, for ground-based detectors the sky coverage is confined to a ring about the celestial pole, so the problem is reduced to a one-dimensional Fourier transform in the azimuthal angle $\phi_a$ as discussed in Ref.~\cite{allen1997}.

Basically, our formalism can be equally applied to space-based GW detectors. For example, in the LISA space mission, the three spacecrafts form an equilateral triangle ABC, as illustrated in Fig.~\ref{fig:lisa}, where the center of mass is located at the origin of the rectangular coordinates XYZ. There are three baseline vectors: AB, BC, and CA, which are also the detector arms. Because LISA is a single instrument sharing all arms and hence the instrumental noises, the method to correlate signals from a detector pair is not applicable in LISA measurements. However, the output of the three detectors can be combined to form two time-delay interferometry (TDI) channels with orthogonal noises~\cite{adams2010}. Hence the correlations between the two TDI channels caused by polarized anisotropic SGWB
can be computed in a similar way as the ground-based detectors. 
A correlation output can be labelled by the pointing angles $(\alpha,\theta,\phi)$ of the spacecraft A. The final product of the space mission would be a SGWB sky map with a sky coverage being the locus of the pointing in the $SO(3)$ group manifold determined by the design of the spacecrafts orbit.

\section{Conclusion}

We have studied the interferometric observation of stochastic gravitational wave background anisotropies. Different from previous works, we have expanded the polarization tensors of gravitational waves in terms of the spin-weighted spherical harmonics. This allows us to avoid tackling complicated tensor calculus and hence provide a systematic way for calculating the antenna pattern functions for all Stokes parameters. Our formalism has explicitly revealed the topology of the data structure that observed correlated signals are defined in the group manifold of the three-dimensional rotation. The correlations between two detectors in the interferometry experiments such as LIGO-Virgo and KAGRA are explicitly constructed in terms of the Wigner D-functions and the Wigner-3j symbols. Our results may be useful for constructing data pipelines to estimate the power spectra of stochastic gravitational wave background anisotropies.

\begin{acknowledgments}
This work was supported in part by the Ministry of Science and Technology (MOST) of Taiwan, R.O.C., under
Grants No. MOST 108-2112-M-032-001- (G.C.L.), No. MOST 108-2112-M-001-008, and No. MOST 109-2112-M-001-003 (K.W.N.).
\end{acknowledgments}

\appendix
\bw
\section{Spin-Weighted Spherical Harmonics}
The explicit form of the spin-weighted spherical harmonics that we use is
\begin{align}
{}_{s}Y_{\ell m}(\theta,\phi) = 
    (-1)^{s+m}
    e^{im\phi}
    \sqrt{\frac{(2\ell+1)}{(4\pi)}\frac{(\ell+m)!(\ell-m)!}{(\ell+s)!(\ell-s)!}}
    \sin^{2\ell}\!\left(\frac{\theta}{2}\right)
    \sum_{r}
    \binom{\ell-s}{r} \binom{\ell+s}{r+s-m}
    (-1)^{\ell-r-s}
    \cot^{2r+s-m}\!\left(\frac{\theta}{2}\right) \,.
\end{align}
\ew
When $s=0$, it reduces to the ordinary spherical harmonics,
\be
Y_{\ell m}(\hat{n})=\sqrt{\frac{(2\ell+1)}{(4\pi)}\frac{(\ell-m)!}{(\ell+m)!}}P_{\ell m}(\cos \theta) e^{i m \phi} \,.
\ee

Spin-weighted spherical harmonics satisfy the orthogonal relation,
\be
    \int_{S^2} \d{\hat{n}}\; {}_{s}Y^*_{\ell m}(\hat{n}){}_{s}Y_{\ell' m'}(\hat{n})
    = \delta_{\ell \ell'} \delta_{m m'} \,,
\ee
and the completeness relation,
\begin{align}
    \sum_{\ell m} {}_{s}Y^*_{\ell m}(\hat{n}){}_{s}Y_{\ell m}(\hat{n}')
    =& \delta(\hat{n}-\hat{n}') \nonumber \\
    =& \delta(\phi-\phi')\delta(\cos\theta-\cos\theta')\,.
\end{align}
Its complex conjugate is
\be
{}_{s}Y^*_{\ell m}(\hat{n}) =  (-1)^{s+m} {}_{-s}Y_{\ell -m}(\hat{n}) \,,
\label{Yconjugate}
\ee
and its parity is given by
\be
{}_{s}Y_{\ell m}(-\hat{n}) \equiv {}_{s}Y_{\ell m}(\pi-\theta,\phi+\pi)=(-1)^{\ell} {}_{-s}Y_{\ell m}(\hat{n}) \,.
\label{Yparity}
\ee

Also, we have the spherical wave expansion:
\be
e^{i\vec{k}\cdot\vec{r}} = 4\pi \suml \summ i^\ell j_\ell(kr) Y_{\ell m}^*(\hat{k}) Y_{\ell m}(\hat{r}) \,,
\ee
where $j_\ell(x)$ is the spherical Bessel function. 


\section{Three-Dimensional Rotation}
\label{sec:3drot}
A three-dimensional rotation can be parameterized by three Euler angles, $\alpha$, $\beta$, and $\gamma$. To rotate a vector in the real space, we may use the Euler matrix $\mathbf{R}(\alpha, \beta, \gamma)$, which can be decomposed into three consecutive rotations around fixed global axes as
\be
\mathbf{R}(\alpha,\beta,\gamma)= \mathbf{R}_Z(\alpha) \mathbf{R}_Y(\beta) \mathbf{R}_Z(\gamma) \,.
\ee
For example,
\be
\vec{Z} = \mathbf{R}(0,\frac{\pi}{2},\frac{\pi}{2})\vec{Y} \,.
\ee
For a tensor constructed by the direct product of a number of vectors, we adopt the convention self-explained by the following example,
\be
    \left[\vec{Z}\otimes\vec{Y}\right]^{ij}
   =\mathbf{R}^{ij}_{kl}(0,\frac{\pi}{2},\frac{\pi}{2})
    \left[\vec{Y}\otimes\vec{X}\right]^{kl} \,.
\ee
Furthermore, we can perform the similar rotation to a function defined on a sphere $S^2$, 
\begin{align}
f^{\mathbf{R}}(\hat{n})
    &=\mathbf{R}(\alpha,\beta,\gamma)f(\hat{n}) \nonumber \\
    &=\langle \hat{n} \vert \mathbf{R}(\alpha,\beta,\gamma) \vert f \rangle \nonumber \\
    &=f(\mathbf{R}^{-1}(\alpha,\beta,\gamma) \hat{n}) \nonumber \\
    &=f(\mathbf{R}(-\beta,-\gamma,-\alpha) \hat{n}) \,.
\end{align}
For instance, 
\begin{align}
f^{\mathbf{R}}(\theta,\phi)
    &=\mathbf{R}(0,0,\delta)f(\theta,\phi) \nonumber \\
    &=f(\theta,\phi-\delta) \,,
\end{align}
in which the function $f$ have been actively rotated about the fixed Z-axis by an angle $\delta$.

If we expand the function in terms of spherical harmonics:
\begin{align}
f(\theta,\phi)= \langle \theta,\phi \vert f \rangle
    &=\sum_{\ell m}
      \langle \theta,\phi \vert \ell,m \rangle
      \langle \ell,m \vert f \rangle \nonumber \\
    &=\sum_{\ell m}Y_{\ell m}(\theta,\phi) f_{\ell m} \,,
\end{align}
the active Euler rotation $\mathbf{R}(\alpha,\beta,\gamma)$ can be acted on either the expansion coefficients or the spherical harmonics:
\begin{align}
    &\langle \theta,\phi \vert \mathbf{R} \vert f \rangle \nonumber \\
    =&
      \sum_{\ell}
      \sum_{ m}
      \sum_{ m'}
      \langle \theta,\phi \vert 
      \ell,m \rangle \langle \ell,m 
      \vert \mathbf{R} \vert
      \ell,m' \rangle \langle \ell,m' 
      \vert f \rangle \nonumber \\
    =&
      \sum_{\ell}
      \sum_{ m}
      \sum_{ m'}
      Y_{\ell m}(\theta,\phi) 
      D^{\ell}_{m m'}(\mathbf{R})
      f_{\ell m'} \nonumber \\ 
    =&
      \sum_{\ell}
      \sum_{ m'}
      Y^{\mathbf{R}}_{\ell m'}(\theta,\phi) 
      f_{\ell m'} \nonumber \\
    =&
      \sum_{\ell}
      \sum_{ m}
      Y_{\ell m}(\theta,\phi) 
      f^{\mathbf{R}}_{\ell m} \,,
\end{align}
where the Wigner-D matrix is defined by
\be
\langle\ell,m\vert\mathbf{R}\vert\ell,m'\rangle=D^{\ell}_{m m'}(\mathbf{R})=D^{\ell}_{m m'}(\alpha,\beta,\gamma) \,,
\ee
which is a $\ell+1$-dimensional irreducible unitary representation of the rotation operator $\mathbf{R}(\alpha,\beta,\gamma)$. The Wigner-D matrix is closely related to the spin-weighted spherical harmonics:
\begin{align}
    \label{eq:DtoSWSH}
    D^{\ell}_{s m}(\alpha,\beta,\gamma)
    = \sqrt{\frac{4\pi}{2\ell+1}} {}_{-s}Y_{\ell m}(-\beta,-\gamma) e^{-i s \alpha} \,.
\end{align}

\begin{figure}[ht!]
\centering
\includegraphics[width=0.5\textwidth]{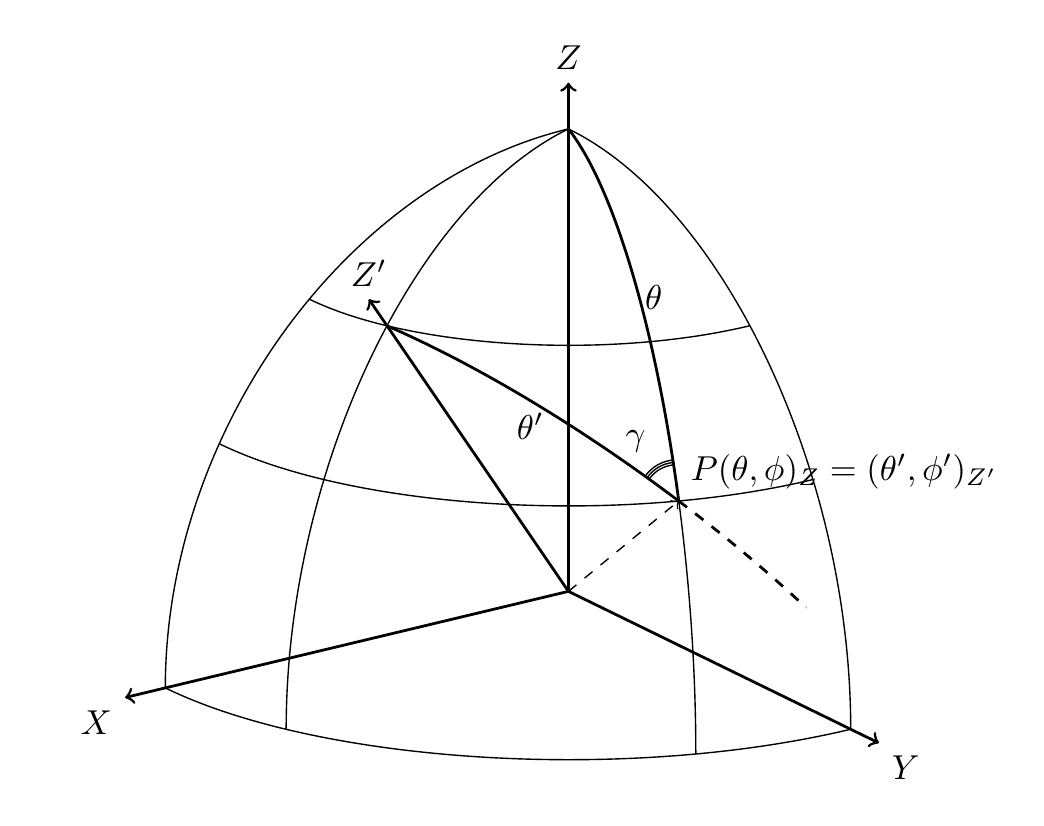}
    \caption[Coordinate]{The angle relates the spin-weighted spherical harmonics in two different coordinates. A point $P$ can be described by both the primed and unprimed coordinates where the $Z'$ and $Z$ indicate their north poles correspondingly. The angle $\gamma$ is the angle between the two great circles that connect the point $P$ to the two north poles.}
\label{fig:swsh}
\end{figure}

Since the multiplication of two successive rotations is still a rotation, the matrix representation should reflect the closure property,
\begin{align}
    D^{\ell}_{-s m}(\mathbf{R}^{-1}_k)
    =
    \sum_{m'}
    D^{\ell}_{-s m'}(\mathbf{R}^{-1}_k \mathbf{R}_\mathbb{D})
    D^{\ell}_{m' m}(\mathbf{R}^{-1}_\mathbb{D}) \,.
\end{align}
Combining Eq.~(\ref{eq:DtoSWSH}), one can relate the spin-weighted spherical harmonics in different coordinates as
\begin{align}
    \label{eq:YYD}
    {}_{s}Y_{\ell m}(\theta_k, \phi_k)
    e^{-is\gamma_k}
    =
    \sum_{m'}
    {}_{s}Y_{\ell m'}(\theta_{k'}, \phi_{k'})
    e^{-is\gamma_{k'}}
    D^{\ell}_{m' m}(\mathbf{R}^{-1}_\mathbb{D}) \,,
\end{align}
in which $\mathbf{R}_{k'}(\phi_{k'}, \theta_{k'}, \gamma_{k'})\equiv(\mathbf{R}^{-1}_\mathbb{D}\mathbf{R}_k)$. 
By choosing $\gamma_k=0$, i.e., making $\mathbf{R}_k = \mathbf{R}_k(\phi_k,\theta_k,0)$, the LHS of Eq.~(\ref{eq:YYD}) becomes the spin-s spherical harmonics in the unprimed coordinates. However, on the RHS, an extra phase factor $e^{-is\gamma_{k'}}$ appears in addition to  the Wigner-D matrix of the coordinate transformation. The angle $\gamma_{k'}(\theta_{k'},\phi_{k'};\mathbf{R}_\mathbb{D})$ is the angle between the two great circles that connect the point $(\theta_{k'},\phi_{k'})$ to the unprimed and the primed north poles as shown in Fig.~\ref{fig:swsh}.  Also see a similar result, the Eq.~(5.4) in Ref.~\cite{ngliu}, derived in a different context.

Here, we summarize some results used in the literature: 
\be
\label{eq:rotYlm}
Y_{\ell m}( \mathbf{R} \hat{n} ) = \sum_{m'} Y_{\ell m'}( \hat{n} ) D^{\ell}_{m' m}( \mathbf{R}^{-1} ) \,,
\ee
\be
\label{eq:rotsYlm}
{}_sY_{\ell m}( \hat{n} ) = \sum_{m'} {}_sY_{\ell m'}( \mathbf{R}^{-1} \hat{n} ) e^{-is\gamma_{n'}} D^{\ell}_{m' m}( \mathbf{R}^{-1} ) \,,
\ee
\begin{align}
    f^{\mathbf{R}^{-1}}_{\ell m} 
    &= \sum_{m'} D^{\ell}_{m m'}( \mathbf{R}^{-1} ) f_{\ell m'} \label{eq:rotflm}
    \nonumber \\
    &= \sum_{m'} D^{*\ell}_{m' m}( \mathbf{R} ) f_{\ell m'} \,.
\end{align}
\bw
     
\section{Multipole Moments of Polarization Tensor}
\ew
\subsection{${}_{ijkl} \mathbb{E}_{\ell m}^{I}$}
The only non-zero coefficients are $\ell=0,2,4$ cases for ${}_{ijkl} \mathbb{E}_{\ell m}^{I}$, which are symmetric under exchanging between $i \leftrightarrow j$, $k \leftrightarrow l$, and ${ij} \leftrightarrow {kl}$. In addition, they satisfy the relation $\mathbb{E}_{\ell -m}^I=(-1)^m \mathbb{E}^{I*}_{\ell m}$.
\begin{align*}
    \frac{16}{15}
    \sqrt{\pi}
  &= {}_{xxxx}\mathbb{E}_{00} 
   = {}_{yyyy}\mathbb{E}_{00}
   = {}_{zzzz}\mathbb{E}_{00}
  \\
    -\frac{8}{15}
    \sqrt{\pi}
  &= {}_{xxyy}\mathbb{E}_{00} 
   = {}_{xxzz}\mathbb{E}_{00} 
   = {}_{yyzz}\mathbb{E}_{00} 
\end{align*}
\begin{align*}
    \frac{16}{21}
    \sqrt{\frac{\pi}{5}}
  &= {}_{xxxx}\mathbb{E}_{20} 
   = {}_{yyyy}\mathbb{E}_{20} 
   = {}_{xxzz}\mathbb{E}_{20}
   = {}_{yyzz}\mathbb{E}_{20}
   \\
    -\frac{32}{21}
    \sqrt{\frac{\pi}{5}}
  &= {}_{zzzz}\mathbb{E}_{20} 
   = {}_{xxyy}\mathbb{E}_{20} 
\end{align*}
\begin{align*}
    \frac{4}{7}
    \sqrt{\frac{2\pi}{15}}
  &= {}_{xxxz}\mathbb{E}_{21} 
   = {}_{xzzz}\mathbb{E}_{21} 
   \\
    -\frac{4i}{7}
    \sqrt{\frac{2\pi}{15}}
  &= {}_{yyyz}\mathbb{E}_{21} 
   = {}_{yzzz}\mathbb{E}_{21} 
   \\
    \frac{2}{7}
    \sqrt{\frac{6\pi}{5}}
  &= {}_{xxyz}\mathbb{E}_{21} 
   = {}_{xyyz}\mathbb{E}_{21} 
\end{align*}
\begin{align*}
    \frac{8}{7}
    \sqrt{\frac{2\pi}{15}}
  &= {}_{xxzz}\mathbb{E}_{22} 
   = {}_{yyyy}\mathbb{E}_{22} 
   \\
    -\frac{8}{7}
    \sqrt{\frac{2\pi}{15}}
  &= {}_{xxxx}\mathbb{E}_{22} 
   = {}_{yyzz}\mathbb{E}_{22} 
   \\
    -\frac{8i}{7}
    \sqrt{\frac{2\pi}{15}}
  &= {}_{xyzz}\mathbb{E}_{22}
   \\
    \frac{4i}{7}
    \sqrt{\frac{2\pi}{15}}
  &= {}_{xxxy}\mathbb{E}_{22} 
   = {}_{xyyy}\mathbb{E}_{22} 
\end{align*}
\begin{align*}
    \frac{2}{35}
    \sqrt{\pi}
  &= {}_{xxxx}\mathbb{E}_{40} 
   = {}_{yyyy}\mathbb{E}_{40} 
   \\
    \frac{2}{105}
    \sqrt{\pi}
  &= {}_{xxyy}\mathbb{E}_{40} 
   \\
    -\frac{8}{105}
    \sqrt{\pi}
  &= {}_{xxzz}\mathbb{E}_{40} 
   = {}_{yyzz}\mathbb{E}_{40} 
   \\
    \frac{16}{105}
    \sqrt{\pi}
  &= {}_{zzzz}\mathbb{E}_{40} 
\end{align*}
\begin{align*}
    \frac{1}{7}
    \sqrt{\frac{\pi}{5}}
  &= {}_{xxxz}\mathbb{E}_{41} 
   \\
    -\frac{i}{7}
    \sqrt{\frac{\pi}{5}}
  &= {}_{yyyz}\mathbb{E}_{41} 
   \\
    \frac{1}{21}
    \sqrt{\frac{\pi}{5}}
  &= {}_{xyyz}\mathbb{E}_{41} 
   \\
    -\frac{i}{21}
    \sqrt{\frac{\pi}{5}}
  &= {}_{xxyz}\mathbb{E}_{41} 
   \\
    -\frac{4}{21}
    \sqrt{\frac{\pi}{5}}
  &= {}_{xzzz}\mathbb{E}_{41} 
   \\
    \frac{4i}{21}
    \sqrt{\frac{\pi}{5}}
  &= {}_{yzzz}\mathbb{E}_{41} 
\end{align*}
\begin{align*}
    \frac{2}{21}
    \sqrt{\frac{2\pi}{5}}
  &= {}_{yyyy}\mathbb{E}_{42} 
   = {}_{xxzz}\mathbb{E}_{42} 
   \\
    -\frac{2}{21}
    \sqrt{\frac{2\pi}{5}}
  &= {}_{xxxx}\mathbb{E}_{42} 
   = {}_{yyzz}\mathbb{E}_{42} 
   \\
    \frac{2i}{21}
    \sqrt{\frac{2\pi}{5}}
  &= {}_{xyzz}\mathbb{E}_{42} 
   \\
    \frac{i}{21}
    \sqrt{\frac{2\pi}{5}}
  &= {}_{xxxy}\mathbb{E}_{42} 
   = {}_{xyyy}\mathbb{E}_{42} 
\end{align*}
\begin{align*}
    \frac{1}{3}
    \sqrt{\frac{\pi}{35}}
  &= {}_{xyyz}\mathbb{E}_{43} 
   \\
    -\frac{1}{3}
    \sqrt{\frac{\pi}{35}}
  &= {}_{xxxz}\mathbb{E}_{43}
   \\
    -\frac{i}{3}
    \sqrt{\frac{\pi}{35}}
  &= {}_{yyyz}\mathbb{E}_{43} 
   \\
    \frac{i}{3}
    \sqrt{\frac{\pi}{35}}
  &= {}_{xxyz}\mathbb{E}_{43} 
\end{align*}
\begin{align*}
    \frac{1}{3}
    \sqrt{\frac{\pi}{35}}
  &= {}_{xyyz}\mathbb{E}_{43} 
   \\
    -\frac{1}{3}
    \sqrt{\frac{\pi}{35}}
  &= {}_{xxxz}\mathbb{E}_{43}
   \\
    -\frac{i}{3}
    \sqrt{\frac{\pi}{35}}
  &= {}_{yyyz}\mathbb{E}_{43} 
   \\
    \frac{i}{3}
    \sqrt{\frac{\pi}{35}}
  &= {}_{xxyz}\mathbb{E}_{43} 
\end{align*}
\begin{align*}
    \frac{1}{3}
    \sqrt{\frac{2\pi}{35}}
  &= {}_{xxxx}\mathbb{E}_{44} 
   = {}_{yyyy}\mathbb{E}_{44} 
   \\
    -\frac{i}{3}
    \sqrt{\frac{2\pi}{35}}
  &= {}_{xxxy}\mathbb{E}_{44}
   \\
    -\frac{1}{3}
    \sqrt{\frac{2\pi}{35}}
  &= {}_{xxyy}\mathbb{E}_{44} 
   \\
    \frac{i}{3}
    \sqrt{\frac{2\pi}{35}}
  &= {}_{xyyy}\mathbb{E}_{44} 
\end{align*}

\subsection{${}_{ijkl} \mathbb{E}_{\ell m}^{V}$}
The only non-zero coefficients are $\ell=1,3$ cases for ${}_{ijkl} \mathbb{E}_{\ell m}^{V}$, which are symmetric under exchanging between $i \leftrightarrow j$ and $k \leftrightarrow l$, and are antisymmetric under exchanging ${ij} \leftrightarrow {kl}$. In addition, they satisfy the relation $\mathbb{E}^V_{\ell -m}=(-1)^{m+1} \mathbb{E}^{V*}_{\ell m}$.  
\begin{align*}
    -\frac{8i}{5}
    \sqrt{\frac{\pi}{3}}
  &= {}_{xxxy}\mathbb{E}_{10} 
   = {}_{xyyy}\mathbb{E}_{10} 
\end{align*}
\begin{align*}
    -\frac{4}{5}
    \sqrt{\frac{2\pi}{3}}
  &= {}_{xxxz}\mathbb{E}_{11} 
   = {}_{xzzz}\mathbb{E}_{11} 
   \\
    \frac{4i}{5}
    \sqrt{\frac{2\pi}{3}}
  &= {}_{yyyz}\mathbb{E}_{11} 
   = {}_{yzzz}\mathbb{E}_{11} 
   \\
    -\frac{2}{5}
    \sqrt{\frac{2\pi}{3}}
  &= {}_{xyyz}\mathbb{E}_{11} 
\end{align*}
\begin{align*}
    -\frac{2i}{5}
    \sqrt{\frac{\pi}{7}}
  &= {}_{xxxy}\mathbb{E}_{30} 
   = {}_{xyyy}\mathbb{E}_{30} 
\end{align*}
\begin{align*}
    -\frac{1}{5}
    \sqrt{\frac{\pi}{21}}
  &= {}_{xxxz}\mathbb{E}_{31} 
   \\
    -i
    \sqrt{\frac{\pi}{21}}
  &= {}_{xxyz}\mathbb{E}_{31} 
   \\
    -\frac{1}{5}
    \sqrt{\frac{3\pi}{7}}
  &= {}_{xyyz}\mathbb{E}_{31} 
   \\
    \frac{4}{5}
    \sqrt{\frac{\pi}{21}}
  &= {}_{xzzz}\mathbb{E}_{31} 
   \\
    \frac{i}{5}
    \sqrt{\frac{\pi}{21}}
  &= {}_{yyyz}\mathbb{E}_{31} 
   \\
    -\frac{4i}{5}
    \sqrt{\frac{\pi}{21}}
  &= {}_{yzzz}\mathbb{E}_{31} 
\end{align*}
\begin{align*}
    i
    \sqrt{\frac{2\pi}{105}}
  &= {}_{xxxy}\mathbb{E}_{32} 
  \\
    2
    \sqrt{\frac{2\pi}{105}}
  &= {}_{xxyy}\mathbb{E}_{32} 
   = {}_{yyzz}\mathbb{E}_{32} 
  \\
    -2
    \sqrt{\frac{2\pi}{105}}
  &= {}_{xxzz}\mathbb{E}_{32} 
  \\
    -i
    \sqrt{\frac{2\pi}{105}}
  &= {}_{xyyy}\mathbb{E}_{32} 
  \\
    2i
    \sqrt{\frac{2\pi}{105}}
  &= {}_{xyzz}\mathbb{E}_{32} 
\end{align*}
\begin{align*}
    \sqrt{\frac{\pi}{35}}
  &= {}_{xxxz}\mathbb{E}_{33} 
  \\
    -i
    \sqrt{\frac{\pi}{35}}
  &= {}_{xxyz}\mathbb{E}_{33} 
  \\
    -
    \sqrt{\frac{\pi}{35}}
  &= {}_{xyyz}\mathbb{E}_{33} 
  \\
    i
    \sqrt{\frac{\pi}{35}}
  &= {}_{yyyz}\mathbb{E}_{33} 
\end{align*}

\subsection{${}_{ijkl} \mathbb{E}_{\ell m}^{Q\pm iU}$}
The only non-zero coefficients are $\ell=4$ cases for ${}_{ijkl} \mathbb{E}_{\ell m}^{Q\pm iU}$, which are symmetric under exchanging between $i \leftrightarrow j$, $k \leftrightarrow l$, and ${ij} \leftrightarrow {kl}$. In addition, they satisfy the relation $\mathbb{E}_{\ell -m}^{Q\pm iU}=(-1)^m \mathbb{E}_{\ell m}^{Q\mp iU*}$. 
Here, specifically we have $\mathbb{E}_{\ell m}^{Q+iU}=\mathbb{E}_{\ell m}^{Q-iU}$.
\begin{align*}
    \sqrt{\frac{2\pi}{35}}
  &= {}_{xxxx}\mathbb{E}_{40} 
   = {}_{yyyy}\mathbb{E}_{40}
  \\
    \frac{1}{3}
    \sqrt{\frac{2\pi}{35}}
 &= {}_{xxyy}\mathbb{E}_{40} 
  \\
    -\frac{4}{3}
    \sqrt{\frac{2\pi}{35}}
 &= {}_{xxzz}\mathbb{E}_{40} 
  = {}_{yyzz}\mathbb{E}_{40}
  \\
    \frac{8}{3}
    \sqrt{\frac{2\pi}{35}}
 &= {}_{zzzz}\mathbb{E}_{40} 
\end{align*}

\begin{align*}
    \sqrt{\frac{\pi}{14}}
  &= {}_{xxxz}\mathbb{E}_{41} 
  \\
    -\frac{i}{3}
    \sqrt{\frac{\pi}{14}}
  &= {}_{xxyz}\mathbb{E}_{41}
  \\
    \frac{1}{3}
    \sqrt{\frac{\pi}{14}}
  &= {}_{xyyz}\mathbb{E}_{41} 
  \\
    -\frac{2}{3}
    \sqrt{\frac{2\pi}{7}}
  &= {}_{xzzz}\mathbb{E}_{41} 
  \\
    -i
    \sqrt{\frac{\pi}{14}}
  &= {}_{yyyz}\mathbb{E}_{41}
  \\
    \frac{2i}{3}
    \sqrt{\frac{\pi}{7}}
  &= {}_{yzzz}\mathbb{E}_{41} 
\end{align*}

\begin{align*}
    \frac{2}{3}
    \sqrt{\frac{\pi}{7}}
  &= {}_{xxzz}\mathbb{E}_{42}
   = {}_{yyyy}\mathbb{E}_{42}
   \\
    -\frac{2}{3}
    \sqrt{\frac{\pi}{7}}
  &= {}_{xxxx}\mathbb{E}_{42} 
   = {}_{yyzz}\mathbb{E}_{42} 
  \\
    \frac{i}{3}
    \sqrt{\frac{\pi}{7}}
  &= {}_{xxxy}\mathbb{E}_{42}
   = {}_{xyyy}\mathbb{E}_{42} 
\end{align*}

\begin{align*}
    \frac{1}{3}
    \sqrt{\frac{\pi}{2}}
  &= {}_{xyyz}\mathbb{E}_{43}
   = {}_{xxxz}\mathbb{E}_{43}
  \\
    \frac{i}{3}
    \sqrt{\frac{\pi}{2}}
  &= {}_{xxyz}\mathbb{E}_{43}
   = {}_{yyyz}\mathbb{E}_{43} 
\end{align*}

\begin{align*}
    \frac{1}{3}
    \sqrt{\pi}
  &= {}_{xxxx}\mathbb{E}_{44}
   = {}_{yyyy}\mathbb{E}_{44}
   \\
    -\frac{1}{3}
    \sqrt{\pi}
  &= {}_{xxyy}\mathbb{E}_{44}
  \\
    \frac{i}{3}
    \sqrt{\pi}
  &= {}_{xyyy}\mathbb{E}_{44}
  \\
    -\frac{i}{3}
    \sqrt{\pi}
  &= {}_{xxxy}\mathbb{E}_{44} 
\end{align*}

\bw
\section{Antenna Pattern Functions}
\label{sec:DE}
In most literature, the inner product between the detector tensor and the polarization basis tensor is referred as the antenna pattern function. 

\subsection{$\mathbb{DE}^I$}
\label{sec:DEI}
For the Stokes-I parts, the only non-vanishing $\mathbb{D}_0(\sigma_1,\sigma_2,\beta) \cdot \mathbb{E}^{I}_{\ell m}$ are 15 components with $\ell=0,2,4$, which satisfy $\mathbb{DE}_{\ell -m}=(-1)^m \mathbb{DE}^*_{\ell m}$.
\begin{align*}
    \mathbb{DE}^I_{00}
    &=
    \frac{4}{5} \sqrt{\pi } \left(\cos ^4\left(\frac{\beta }{2}\right) \cos (2 \sigma_1- 2 \sigma_2 )+\sin ^4\left(\frac{\beta }{2}\right) \cos (2 \sigma_1+ 2 \sigma_2 )\right)
    \\
    \mathbb{DE}^I_{20}
    &=
    \frac{8}{7} \sqrt{\frac{\pi }{5}} \left(\cos ^4\left(\frac{\beta }{2}\right) \cos (2 \sigma_1- 2 \sigma_2 )+\sin ^4\left(\frac{\beta }{2}\right) \cos (2 \sigma_1+ 2 \sigma_2 )\right)
    \\
    \mathbb{DE}^I_{21}
    &=
    -\frac{2}{7} \sqrt{\frac{6 \pi }{5}} e^{-2 i \sigma_1} \sin (\beta ) (\cos (\beta ) \cos (2 \sigma_2)+i \sin (2 \sigma_2))
    \\
    \mathbb{DE}^I_{22}
    &=
    \frac{2}{7} \sqrt{\frac{6 \pi }{5}} e^{-2 i \sigma_1} \sin ^2(\beta ) \cos (2 \sigma_2)
    \\
    \mathbb{DE}^I_{40}
    &=
    \frac{2}{105} \sqrt{\pi } \left(\cos ^4\left(\frac{\beta }{2}\right) \cos (2 \sigma_1 - 2 \sigma_2)+\sin ^4\left(\frac{\beta }{2}\right) \cos (2 \sigma_1 + 2 \sigma_2 )\right)
    \\
    \mathbb{DE}^I_{41}
    &=
    -\frac{1}{21} \sqrt{\frac{\pi }{5}} e^{-2 i \sigma_1} \sin (\beta ) (\cos (\beta ) \cos (2 \sigma_2)+i \sin (2 \sigma_2))
    \\
    \mathbb{DE}^I_{42}
    &=
    \frac{1}{7} \sqrt{\frac{\pi }{10}} e^{-2 i \sigma_1} \sin ^2(\beta ) \cos (2 \sigma_2)
    \\
    \mathbb{DE}^I_{43}
    &=
    \frac{1}{3} \sqrt{\frac{\pi }{35}} e^{-2 i \sigma_1} \sin (\beta ) (\cos (\beta ) \cos (2 \sigma_2)-i \sin (2 \sigma_2))
    \\
    \mathbb{DE}^I_{44}
    &=
    \frac{1}{6} \sqrt{\frac{\pi }{70}} e^{-2 i \sigma_1} ((\cos (2 \beta )+3) \cos (2 \sigma_2)-4 i \cos (\beta ) \sin (2 \sigma_2))
\end{align*}

\subsection{$\mathbb{DE}^V$}
\label{sec:DEV}
For the Stokes-V parts that correspond to the circular polarized signal, the only non-vanishing $\mathbb{D}_0(\sigma_1,\sigma_2,\beta) \cdot \mathbb{E}^{V}_{\ell m}$ are 10 components with $\ell=1,3$, which satisfy $\mathbb{DE}_{\ell -m}=(-1)^{m+1} \mathbb{DE}^*_{\ell m}$.

\begin{align*}
    \mathbb{DE}^V_{10}
    &=
    \frac{8}{5} i \sqrt{\frac{\pi }{3}} \left(\cos ^4\left(\frac{\beta }{2}\right) \sin (2 \sigma_1 - 2 \sigma_2 )+\sin ^4\left(\frac{\beta }{2}\right) \sin (2 \sigma_1 + 2\sigma_2 )\right)
    \\
    \mathbb{DE}^V_{11}
    &=
    \frac{2}{5} \sqrt{\frac{2 \pi }{3}} e^{-2 i \sigma_1} \sin (\beta ) (\cos (\beta ) \cos (2 \sigma_2)+i \sin (2 \sigma_2))
    \\
    \mathbb{DE}^V_{30}
    &=
    \frac{2}{5} i \sqrt{\frac{\pi }{7}} \left(\cos ^4\left(\frac{\beta }{2}\right) \sin (2 \sigma_1 - 2 \sigma_2 )+\sin ^4\left(\frac{\beta }{2}\right) \sin (2 \sigma_1 + 2 \sigma_2 )\right)
    \\
    \mathbb{DE}^V_{31}
    &=
    \frac{1}{5} \sqrt{\frac{3 \pi }{7}} e^{-2 i \sigma_1} \sin (\beta ) (\cos (\beta ) \cos (2 \sigma_2)+i \sin (2 \sigma_2))
    \\
    \mathbb{DE}^V_{32}
    &=
    -\sqrt{\frac{3 \pi }{70}} e^{-2 i \sigma_1} \sin ^2(\beta ) \cos (2 \sigma_2)
    \\
    \mathbb{DE}^V_{33}
    &=
    -\sqrt{\frac{\pi }{35}} e^{-2 i \sigma_1} \sin (\beta ) (\cos (\beta ) \cos (2 \sigma_2)-i \sin (2 \sigma_2))
\end{align*}

\subsection{$\mathbb{DE}^{Q\pm iU}$}
For the linear polarized signal, the only non-vanishing $\mathbb{D}_0(\sigma_1,\sigma_2,\beta) \cdot \mathbb{E}^{Q\pm i U}_{\ell m}$ are 9 components with $\ell=4$, which satisfy $\mathbb{DE}_{\ell -m}=(-1)^m \mathbb{DE}^*_{\ell m}$.
\begin{align*}
    \mathbb{DE}^{Q\pm iU}_{40}
    &=
    \frac{1}{3} \sqrt{\frac{2 \pi }{35}} \left(\cos ^4\left(\frac{\beta }{2}\right) \cos (2 \sigma_1- 2 \sigma_2 )+\sin ^4\left(\frac{\beta }{2}\right) \cos (2 \sigma_1+ 2 \sigma_2)\right)
    \\
    \mathbb{DE}^{Q\pm iU}_{41}
    &=
    -\frac{1}{3} \sqrt{\frac{\pi }{14}} e^{-2 i \sigma_1} \sin (\beta ) (\cos (\beta ) \cos (2 \sigma_2)+i \sin (2 \sigma_2))
    \\
    \mathbb{DE}^{Q\pm iU}_{42}
    &=
    \frac{1}{2} \sqrt{\frac{\pi }{7}} e^{-2 i \sigma_1} \sin ^2(\beta ) \cos (2 \sigma_2)
    \\
    \mathbb{DE}^{Q\pm iU}_{43}
    &=
    \frac{1}{3} \sqrt{\frac{\pi }{2}} e^{-2 i \sigma_1} \sin (\beta ) (\cos (\beta ) \cos (2 \sigma_2)-i \sin (2 \sigma_2))
    \\
    \mathbb{DE}^{Q\pm iU}_{44}
    &=
    \frac{1}{12} \sqrt{\pi } e^{-2 i \sigma_1} ((\cos (2 \beta )+3) \cos (2 \sigma_2)-4 i \cos (\beta ) \sin (2 \sigma_2))
\end{align*}

\ew

\newcommand{\Authname}[2]{#2 #1} 
\newcommand{\etal}{{\it et al.}}
\newcommand{\LSC}{\Authname{Abbott}{B.~P.} \etal\, (LIGO Scientific Collaboration)}
\newcommand{\LVC}{\Authname{Abbott}{B.~P.} \etal\, (LIGO Scientific and Virgo Collaboration)}
\newcommand{\LVK}{\Authname{Abbott}{B.~P.} \etal\, (LIGO Scientific, Virgo and KAGRA Collaboration)}

\newcommand{\Title}[1]{}               

\newcommand{\arxiv}[1]{\href{http://arxiv.org/abs/#1}{{arXiv:}#1}}
\newcommand{\PRD}[3]{\href{https://doi.org/10.1103/PhysRevD.#1.#2}{{Phys. Rev. D} {\bf #1}, #2 (#3)}}
\newcommand{\PRDR}[3]{\href{https://doi.org/10.1103/PhysRevD.#1.#2}{{Phys. Rev. D} {\bf #1}, #2(R) (#3)}}
\newcommand{\PRL}[3]{\href{https://doi.org/10.1103/PhysRevLett.#1.#2}{{Phys. Rev. Lett.} {\bf #1}, #2 (#3)}}

\newcommand{\MNRAS}[4]{\href{https://doi.org/10.1093/mnras/#1.#4.#2}{{Mon. Not. R. Astron. Soc.} {\bf #1}, #2 (#3)}}
\newcommand{\CQGii}[5]{\href{https://doi.org/10.1088/0264-9381/#1/#4/#5}{{Class. Quant. Grav.} {\bf #1}, #2 (#3)}}
\newcommand{\CQG}[4]{\href{https://doi.org/10.1088/1361-6382/#4}{{Class. Quant. Grav.} {\bf #1}, #2 (#3)}}
\newcommand{\JCAP}[3]{\href{https://doi.org/10.1088/1475-7516/#3/#1/#2}{{J. Cosmol. Astropart. Phys.} #1 (#3) #2}}

\newcommand{\LRR}[4]{\href{https://doi.org/10.1007/#4}{{Liv. Rev. Rel.} {\bf #1}, #2 (#3)}}
\newcommand{\ApJ}[4]{\href{https://doi.org/#4}{{Astrophys. J.} {\bf #1}, #2 (#3)}}


\raggedright


\begin{thebibliography}{99}

\bibitem{ligo}
    \LVC,
    \PRL{116}{061102}{2016}.

\bibitem{ligo2019}
    \LVC,
    \CQG{37}{055002}{2020}{ab685e}.

\bibitem{et2010}
    \Authname{Punturo}{M.} \etal,
    \Title{The Einstein Telescope: A third-generation gravitational wave observatory}
    \CQGii{27}{194002}{2010}{19}{194002}.

\bibitem{ce2019}
    \Authname{Reitze}{D.} \etal, 
    \Title{Cosmic Explorer: The U.S. Contribution to Gravitational-Wave Astronomy beyond LIGO}
    Bull. Am. Astron. Soc. \textbf{51}, no.7, 035 (2019)
    \arxiv{1907.04833}.

\bibitem{lisa2017}
    \Authname{Amaro-Seoane}{P.} \etal\, (LISA),
    \Title{Laser Interferometer Space Antenna}
    \arxiv{1702.00786}.

\bibitem{decigo2011}
    \Authname{Kawamura}{S.} \etal,
    \Title{The Japanese space gravitational wave antenna: DECIGO}
    \CQGii{28}{094011}{2011}{9}{094011}.

\bibitem{taiji2017}
    \Authname{Hu}{W.~R.} and 
    \Authname{Wu}{Y.~L.},
    \Title{The Taiji Program in Space for gravitational wave physics and the nature of gravity}
    \href{https://doi.org/10.1093/nsr/nwx116}{{Natl. Sci. Rev.} {\bf 4}, no.5, 685-686 (2017)}.

\bibitem{tianqin2016}
    \Authname{Luo}{J.} \etal\, (TianQin),
    \Title{TianQin: a space-borne gravitational wave detector}
    \CQGii{33}{035010}{2016}{3}{035010}.

\bibitem{ska2015}
    \Authname{Janssen}{G.} \etal,
    \Title{Gravitational wave astronomy with the SKA}
    \href{https://doi.org/10.22323/1.215.0037}{{PoS} {\bf AASKA14}, 037, (2015)}.

\bibitem{ligo2050}
    For examples, see M. A. Sedda {\it et al.}, \arxiv{1908.11375}; V. Baibhav {\it et al.}, \arxiv{1908.11390}; J. Baker {\it et al.}, \arxiv{1908.11410}.

\bibitem{romano}
    For a review, see J. D. Romano, \arxiv{1909.00269}.

\bibitem{alexander}
\Authname{Alexander}{S.~H.~S.},
    \Authname{Peskin}{M.~E.}, and 
    \Authname{Sheikh-Jabbari}{M.~M.},
    \Title{Leptogenesis from gravity waves in models of inflation}
    \PRL{96}{081301}{2006}.
    
\bibitem{satoh}
\Authname{Satoh}{M.},
    \Authname{Kanno}{S.}, and 
    \Authname{Soda}{J.},
    \Title{Circular polarization of primordial gravitational waves in string-inspired inflationary cosmology}
    \PRD{77}{023526}{2008}.
    
\bibitem{sorbo}
\Authname{Sorbo}{L.},
    \Title{Parity violation in the Cosmic Microwave Background from a pseudoscalar inflaton}
    \JCAP{06}{003}{2011}.

\bibitem{crowder2013}
    \Authname{Crowder}{S.~G.},
    \Authname{Namba}{R.},
    \Authname{Mandic}{V.},
    \Authname{Mukohyama}{S.}, and 
    \Authname{Peloso}{M.},
    \Title{Measurement of parity violation in the early universe using gravitational-wave detectors}
     \href{https://doi.org/10.1016/j.physletb.2013.08.077}{Phys. Lett. B {\bf 726}, 66 (2013)}.
   
\bibitem{cusin}
\Authname{Cusin}{G.},
    \Authname{Durrer}{R.}, and 
    \Authname{Ferreira}{P.~G.},
    \Title{Polarization of a stochastic gravitational wave background through diffusion by massive structures}
    \PRD{99}{023534}{2019}.
    
\bibitem{bartolo}
    \Authname{Bartolo}{N.},
    \Authname{Bertacca}{D.},
    \Authname{ Matarrese}{S.},
    \Authname{Peloso}{M.},
    \Authname{Ricciardone}{A.},
    \Authname{Riotto}{A.}, and 
    \Authname{Tasinato}{G.},
    \Title{Characterizing the cosmological gravitational wave background: Anisotropies and non-Gaussianity}
    \PRD{102}{023527}{2020}.
   
\bibitem{pitrou}
\Authname{Pitrou}{C.},
    \Authname{Cusin}{G.}, and 
    \Authname{Uzan}{J.-P.},
    \Title{Unified view of anisotropies in the astrophysical gravitational-wave background}
    \PRDR{101}{081301}{2020}.

\bibitem{michelson1987} 
    \Authname{Michelson}{P.~F.},
    \Title{On detecting stochastic background gravitational radiation with terrestrial detectors}
    \MNRAS{227}{933}{1987}{4}.

\bibitem{christensen1992}
    \Authname{Christensen}{N.},
    \Title{Measuring the stochastic gravitational-radiation background with laser-interferometric antennas}
    \PRD{46}{5250}{1992}.

\bibitem{flanagan1993}
    \Authname{Flanagan}{E.~E.},
    \Title{Sensitivity of the Laser Interferometer Gravitational Wave Observatory to a stochastic background, and its dependence on the detector orientations}
    \PRD{48}{2389}{1993}.

\bibitem{allen1997}
    \Authname{Allen}{B.} and \Authname{Ottewill}{A.~C.},
    \Title{Detection of anisotropies in the gravitational-wave stochastic background}
    \PRD{56}{545}{1997}.

\bibitem{allen1999}
    \Authname{Allen}{B.} and \Authname{Romano}{J.~D.},
    \Title{Detecting a stochastic background of gravitational radiation: Signal processing strategies and sensitivities}
    \PRD{59}{102001}{1999}.

\bibitem{cornish2001} 
    \Authname{Cornish}{N.~J.},
    \Title{Mapping the gravitational-wave background}
    \CQGii{18}{4277}{2001}{20}{307}.

\bibitem{seto2006} 
    \Authname{Seto}{N.},
    \Title{Prospects for Direct Detection of the Circular Polarization of the Gravitational-Wave Background}
     \PRL{97}{151101}{2006}.

\bibitem{seto2007} 
    \Authname{Seto}{N.},
    \Title{Quest for circular polarization of a gravitational wave background and orbits of laser interferometers in space}
    \PRDR{75}{061302}{2007}.

\bibitem{seto2008} 
    \Authname{Seto}{N.} and \Authname{Taruya}{A.},
    \Title{Polarization analysis of gravitational-wave backgrounds from the correlation signals of ground-based interferometers: Measuring a circular-polarization mode}
    \PRD{77}{103001}{2008}.

\bibitem{thrane2009}
    \Authname{Thrane}{E.},
    \Authname{Ballmer}{S.},
    \Authname{Romano}{J.~D.},
    \Authname{Mitra}{S.},
    \Authname{Talukder}{D.},
    \Authname{Bose}{S.}, and 
    \Authname{Mandic}{V.},
    \Title{Probing the anisotropies of a stochastic gravitational-wave background using a network of ground-based laser interferometers}
    \PRD{80}{122002}{2009}.
    
\bibitem{romano2017}
    \Authname{Romano}{J.~D.} and
    \Authname{Cornish}{N.~J.},
    \LRR{20}{2}{2017}{s41114-017-0004-1}

\bibitem{ligo2019_O2iso} 
    \LVC,
    \Title{Search for the isotropic stochastic background using data from Advanced LIGO's second observing run}
    \PRD{100}{061101}{2019}.

\bibitem{ligo2019_O2aniso}
    \LVC,
    \Title{Directional limits on persistent gravitational waves using data from Advanced LIGOÕs first two observing runs}
    \PRD{100}{062001}{2019}.


\bibitem{book:BornAndWolf} 
    \Authname{Born}{M.} and \Authname{Wolf}{E.},
    Principles of Optics, 6th Ed.
    (Pergamon Press, New York, 1980).

    

\bibitem{code:lalsuite} 
    LIGO Scientific Collaboration,
    LIGO Algorithm Library-LALSuite, Free Software (GPL),
    2018, \href{https://doi.org/10.7935/GT1W-FZ16}{https://doi.org/10.7935/GT1W-FZ16}


\bibitem{book:Varshalovich} 
    \Authname{Varshalovich}{D.~A.}, \Authname{Moskalev}{A.~N.}, and \Authname{Khersonskii}{V.~K.},
    Quantum Theory of Angular Momentum
    (World Scientific, Singapore, 1988).

\bibitem{challinor2000}
 \Authname{Challinor}{A.},
    \Authname{Fosalba}{P.},
    \Authname{Mortlock}{D.},
    \Authname{Ashdown}{M.},
    \Authname{Wandelt}{B.}, and 
    \Authname{G\'{o}rski}{K.},
    \Title{All-sky convolution for polarimetry experiments}
    \PRD{62}{123002}{2000}.

\bibitem{wandelt2001}
 \Authname{Wandelt}{B.} and 
    \Authname{G\'{o}rski}{K.},
\Title{Fast convolution on the sphere}
    \PRD{63}{123002}{2001}.

\bibitem{adams2010}
 \Authname{Adams}{M.~R.} and 
    \Authname{Cornish}{N.~J.},
\Title{Discriminating between a stochastic gravitational wave background and instrument noise}
    \PRD{82}{022002}{2010}.

\bibitem{ngliu}
    K.-W. Ng and G.-C. Liu, \href{https://doi.org/10.1142/S0218271899000079}{Int. J. Mod. Phys. D {\bf 08}, 61 (1999)}.

\end{thebibliography}
\end{document}